\documentclass[journal=jctcce,manuscript=article]{achemso}

\usepackage{graphicx}% Include figure files
\usepackage{dcolumn}% Align table columns on decimal point
\usepackage{bm}% bold math
\usepackage{glossaries}
\usepackage[version=4]{mhchem}
\usepackage{comment}
\usepackage{algpseudocode}
\usepackage{algorithm}
\usepackage{siunitx}
\usepackage{multirow}
\usepackage{amsmath}
\usepackage{amssymb}
\usepackage{dsfont}
\newacronym{paw}{PAW}{projector augmented wave}
\newacronym{pm}{PM}{Pipek-Mezey}
\newacronym{ibo}{IBO}{intrinsic bond orbital}
\newacronym{iao}{IAO}{intrinsic atomic orbital}
\newacronym{fb}{FB}{Foster-Boys}
\newacronym{er}{ER}{Edmiston-Ruedenberg}
\newacronym{vn}{VN}{von-Niessen}
\newacronym{cc}{CC}{coupled cluster}
\newacronym{hf}{HF}{Hartree-Fock}
\newacronym{dft}{DFT}{density functional theory}
\newacronym{dmc}{DMC}{Diffusion Monte Carlo}
\newacronym{ppl}{ppl}{particle-particle ladder}
\newacronym{rpa}{RPA}{random phase approximation}
\newacronym{ccsd}{CCSD}{coupled cluster with single and double particle–hole excitation operators}
\newacronym{ccsdpt}{CCSD(T)}{coupled cluster with single, double and perturbative triple particle–hole excitation operators}
\newacronym{bvk}{BvK}{Born-von Karman}
\newacronym{bfgs}{BFGS}{Broyden–Fletcher–Goldfarb–Shanno}
\newacronym{lbfgs}{L-BFGS}{limited-memory BFGS}
\newacronym{cg}{CG}{Conjugate-Gradient}
\newacronym{sa}{SA}{Steepest Ascent}
\newacronym{diis}{DIIS}{Direct Inversion in the Iterative Subspace}
\newacronym{vasp}{VASP}{Vienna Ab initio Simulation Package}
\newacronym{pr}{PR}{Polak-Ribière}
\newacronym{bz}{BZ}{Brillouin zone}

\newcommand{\D}{\text d}
\newcommand{\ImI}{\text i}
\newcommand{\EuE}{\text e}

\author{Benjamin W\"ockinger}
\author{Alexander Rumpf}
\author{Tobias Sch\"afer}
\email{tobias.schaefer@tuwien.ac.at}
\affiliation{Institute for Theoretical Physics, TU Wien, Wiedner Hauptstraße 8-10/136, A-1040 Vienna, Austria}

\title{Exploring the Convergence and Properties of Intrinsic Bond Orbitals in Solids}

\begin{document}

\begin{tocentry}
\includegraphics[width=1.0\linewidth]{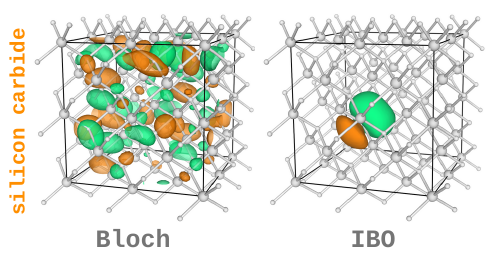}
\end{tocentry}

\begin{abstract}
We present a study of the construction and spatial properties of localized Wannier orbitals in large supercells of insulating solids using plane waves as the underlying basis. 
The \gls{pm} functional in combination with \glspl{iao} as projectors is employed, resulting in so-called  \glspl{ibo}.
Independent of the bonding type and band gap, a correlation between orbital spreads and geometric properties is observed.
As a result, comparable sparsity patterns of the Hartree-Fock exchange matrix are found across all considered bulk 3D materials, exhibiting covalent bonds, polar covalent bonds, and ionic bonds.
Recognizing the considerable computational effort required to construct localized Wannier orbitals for large periodic simulation cells, we address the performance and scaling of different solvers for the localization problem.
This includes the \gls{bfgs}, \gls{cg}, \gls{sa} as well as the \gls{diis} method.
Each algorithm performs a Riemannian optimization under unitary matrix constraint, efficiently reaching the optimum in the ``curved parameter space'' on geodesics.
We hereby complement the quantum chemistry and materials science literature with an introduction to this topic along with key references.
%Also the \gls{diis} technique is employed.
The solvers have been implemented both within the \gls{vasp} and as a standalone open-source software package.
Furthermore, we observe that the construction of Wannier orbitals for supercells of metal oxides presents a significant challenge, requiring approximately one order of magnitude more iteration steps than other systems studied. 
\end{abstract}

%%%%%%%%%%%%%%%%%%%%%%%%%%%%%%%%%%%%%%%%%%%%%%%%%%%%%%
%%%%%%%%%%%%%%%%%%%%%%%%%%%%%%%%%%%%%%%%%%%%%%%%%%%%%%
%      INTRODUCTION
%%%%%%%%%%%%%%%%%%%%%%%%%%%%%%%%%%%%%%%%%%%%%%%%%%%%%%
%%%%%%%%%%%%%%%%%%%%%%%%%%%%%%%%%%%%%%%%%%%%%%%%%%%%%%
\section{Introduction}

Localized orbitals are a useful tool in quantum chemistry and materials physics.
They serve a variety of purposes, for example, the analysis of chemical bonds in tune with chemical intuition~\cite{Edmiston1963,Knizia2013}, the investigation of electron transfer processes~\cite{Knizia2015}, the calculation of electron-phonon interactions~\cite{Engel2020}, or the development of efficient many-electron correlation algorithms by introducing sparsity in electron repulsion integrals~\cite{Voloshina2011,Usvyat2018,Kubas2016,Schafer2021a,Schafer2021b,Lau2021,Ye2024}.

Known as localized Wannier orbitals in solid-state physics and localized molecular orbitals in quantum chemistry, they are usually derived from delocalized one-electron mean-field orbitals through rotations, achieved by a unitary matrix, resulting in spatial confinement.
Various definitions have been proposed for determining this unitary matrix, with several implementations available for periodic systems.
Spatial confinement can be achieved by minimizing the orbital spread, a technique known as \gls{fb} localization~\cite{Foster1960}. 
Alternatively, maximizing electronic self-repulsion, termed \gls{er} localization~\cite{Edmiston1963}, or maximizing self-overlap, known as \gls{vn} localization~\cite{Niessen1973}, can be employed. 
Another approach, \acrfull{pm} localization~\cite{Pipek1989}, utilizes atomic partial charges as the localization measure. 
While these methods require iterative optimization, single-shot localization techniques also exist for solids~\cite{Engel2020,Schafer2021a,Ozaki2023}.

Early implementations for periodic boundary conditions primarily focused on the \gls{fb} localization scheme, with applications for plane-wave basis sets~\cite{Marzari1997,Marzari2012} and atom-centered basis functions~\cite{Zicovic2001}. 
While Riemannian optimization strategies for determining the optimal unitary transformation matrix were applied to molecules by Lehtola et al. in Ref. ~\citenum{lehtola2013unitary}, the \gls{pm} localization technique was adapted to periodic systems using a Riemannian optimization approach by J\'{o}nsson et al. in Ref.~\citenum{Jonsson2017}.
They introduced the notion of generalized \gls{pm} Wannier orbitals by employing various partial charge estimates for the \gls{pm} functional.
Building on the work by J\'{o}nsson et al., subsequent implementations of generalized \gls{pm} Wannier orbitals for solids were reported in Refs.~\citenum{Schafer2021b,Clement2021,Schreder2024,zhu2024wannier}.
The \acrfullpl{ibo} discussed in this work can similarly be understood as a form of generalized PM Wannier functions.

A central challenge in constructing these localized Wannier orbitals lies in efficiently determining the optimal unitary transformation. 
This process can be considered as a Riemannian optimization problem under unitary constraints, but the performance of different algorithms within this framework is not fully clear. Previous work, such as that by Clement et al.~\cite{Clement2021}, suggested the superiority of the \gls{lbfgs} over the \acrfull{cg} solver. 
%However, our investigations within the Riemannian optimization context reveal a different picture, demonstrating that both solvers exhibit comparable performance.
%A potentially crucial distinction lies in the modeling of a solid: while Clement et al. focused on unit cells with k-point sampling in combination with atom-centered basis sets, our work uses the plane wave basis and centers on large supercells with $\bm \Gamma$-only sampling of the \gls{bz}. 
%This difference in the representation of a solid makes the observed discrepancy in solver performance particularly intriguing.
Our investigations within the Riemannian optimization context, however, reveal a different picture, demonstrating that both solvers exhibit comparable performance. 
We attribute this discrepancy to fundamental differences in the computational setup compared to Clement et al.'s approach, which we discuss within our study.
Furthermore, the scalability of Wannier orbital construction with respect to system size remains a significant challenge, especially for applications targeting realistic models of surfaces and defects, which necessitate large simulation cells.
We address the critical question of how the number of iterations required for convergence scales with the number of atoms, providing crucial insights for the application of localized orbitals to increasingly complex materials.
Additionally, we assess whether the \acrfull{diis} technique~\cite{Pulay1980,Pulay1982} can accelerate the convergence of the iterative optimization.

Finally, a key objective of our work is to leverage localized orbitals to introduce sparsity into Coulomb integrals, aiming to mitigate the computational bottleneck of wavefunction based methods.
A prevailing concern has been the potential impact of small band gaps on the sparsity of electron repulsion integrals, which could hinder the effectiveness of local correlation approaches.
Here we investigate the sparsity of the Fock exchange matrix and demonstrate that, for the semiconductors considered, the sparsity is remarkably robust and largely unaffected by the band gap.
%This is a highly encouraging result, reinforcing the potential of localized orbitals for efficient and accurate electronic structure calculations in diverse materials.

The paper is divided into two main parts.
The first main part starts with Sec. \ref{sec:theory} and discusses the theory, implementation, and performance of different numerical solvers to numerically construct \glspl{ibo}.
This part also aims to complement the existing literature by providing a pedagogical mathematical introduction to the topic of  Riemannian optimization under unitary matrix constraint, along with key references essential for those starting in this area.
The second main part starts with Sec. \ref{sec:prop} where we report spatial properties of \glspl{ibo},  an analysis of the sparsity of the Fock exchange matrix, and trends across the considered materials.

%%%%%%%%%%%%%%%%%%%%%%%%%%%%%%%%%%%%%%%%%%%%%%%%%%%%%%
%%%%%%%%%%%%%%%%%%%%%%%%%%%%%%%%%%%%%%%%%%%%%%%%%%%%%%
%      PART I
%%%%%%%%%%%%%%%%%%%%%%%%%%%%%%%%%%%%%%%%%%%%%%%%%%%%%%
%%%%%%%%%%%%%%%%%%%%%%%%%%%%%%%%%%%%%%%%%%%%%%%%%%%%%%
\part[I]{}

%%%%%%%%%%%%%%%%%%%%%%%%%%%%%%%%%%%%%%%%%%%%%%%%%%%%%%
%%%%%%%%%%%%%%%%%%%%%%%%%%%%%%%%%%%%%%%%%%%%%%%%%%%%%%
%      Theory
%%%%%%%%%%%%%%%%%%%%%%%%%%%%%%%%%%%%%%%%%%%%%%%%%%%%%%
%%%%%%%%%%%%%%%%%%%%%%%%%%%%%%%%%%%%%%%%%%%%%%%%%%%%%%
\section{Theory} \label{sec:theory}

\subsection{Intrinsic Bond Orbitals}

\begin{figure}
    \centering
    \includegraphics[width=0.8\linewidth]{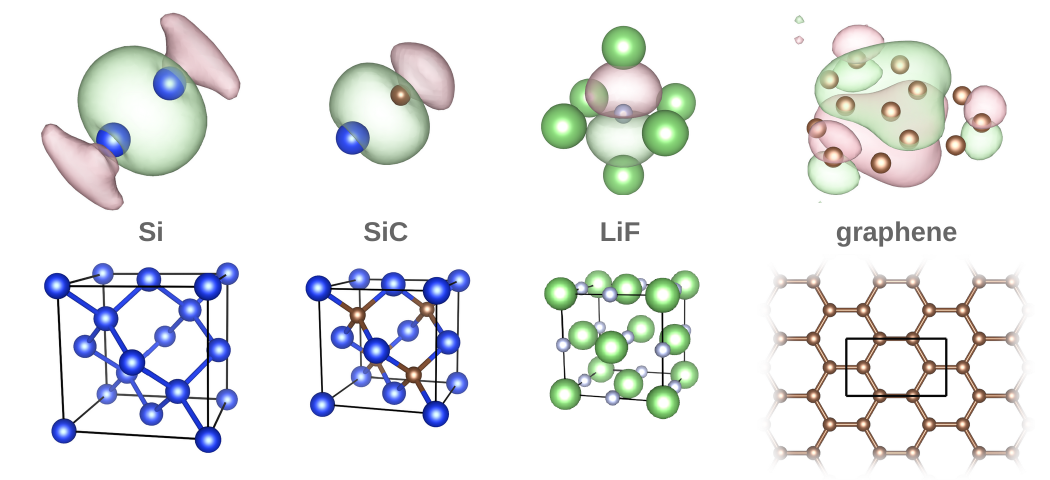}
    \caption{Visual representation of \acrfullpl{ibo} in a selection of materials. The top row shows an \gls{ibo} only with those sites of the periodic structure it connects, indicating the bond. The bottom row shows the conventional unit cell of the corresponding material. All pictures were made with VESTA~\cite{Momma2011}, using an isosurface level of 5.0 for the orbitals.}
    \label{fig:ibos}
\end{figure}

In solids, \acrfullpl{ibo}, $|\mathcal W_{\bm R j}\rangle$, can be defined as generalized Wannier orbitals~\cite{Schafer2021b,Marzari2012}. They are constructed as superpositions of Bloch orbitals, $|\chi_{j\bm k}\rangle$, obtained from prior mean-field calculations such as \gls{hf} or Kohn-Sham \gls{dft},
\begin{equation}
|\mathcal W_{\bm R j}\rangle = \frac{1}{V_\text{BZ}} \int_\text{BZ} \D^3 k \; \EuE^{-\ImI \bm k \bm R} \sum_i^{N_{\text{occ}}} u^{(\bm k)}_{ij} \, |\chi_{j\bm k}\rangle \;.\label{eq:Wannier}
\end{equation}
Here, $u^{(\bm k)}_{ij}$ is a unitary matrix at each k-point $\bm k$, and $V_\text{BZ}$ represents the volume of the \gls{bz}.

In this work, all calculations are based on \gls{hf} orbitals obtained from the plane-wave based \gls{vasp} \cite{Kresse1993,Kresse1996a,Kresse1996b}.
Supercells are considered using a $\bm \Gamma$-only sampling of the \gls{bz}, reducing Eq. (\ref{eq:Wannier}) to 
\begin{equation}
|\mathcal W_{j}\rangle = \sum_i^{N_{\text{occ}}} u_{ij} \, |\chi_{j}\rangle \;.
\end{equation}
The matrix $u_{ij}$ is optimized to maximize (minimize) a localization functional $\mathcal{L}$, which defines the localized Wannier orbitals.
Various localization functionals exist in the literature, such as \gls{fb}~\cite{Foster1960}, \gls{er}~\cite{Edmiston1963}, \gls{vn}~\cite{Niessen1973}, and \acrfull{pm}~\cite{Pipek1989}. 
Our Riemannian optimization algorithm~\cite{lucon} described in Sec. \ref{sec:optimization} is suited for any cost functional, allowing us to compare the case of \gls{pm}, \gls{fb}, and \gls{vn}.
Since we employ $\bm \Gamma$-point-only sampling, the unitary matrices here are in fact real and orthogonal matrices.

\Acrlongpl{ibo} were introduced by Knizia~\cite{Knizia2013} and are the result of maximizing the \gls{pm} functional,
\begin{equation}
\mathcal L^\text{PM}[\{u_{ij}\}] = \sum_i^{N_{\text{occ}}} \sum_A^{N_\text{atoms}} | \, \langle  \mathcal W_i | \bm P_A | \mathcal W_i \rangle \, |^2 \;,
\end{equation}
where $\bm P_A = \sum_{\mu \in A} | \mu\rangle\langle \mu |$ are projectors onto a certain set of atom-centered functions $|\mu\rangle$, also known as \acrfullpl{iao}. 
This choice provides an unbiased measure of atomic partial charges and addresses the well-known basis set dependence associated with Mulliken populations.
While alternative partial charge estimates have been proposed to address this issue for molecules~\cite{Lehtola2014} and also for periodic systems,~\cite{Jonsson2017} \gls{iao}-based charges also independently demonstrated their ability to accurately characterize bonding even in nontrivial transition structures of chemical reactions.~\cite{Knizia2015}
Figure \ref{fig:ibos} illustrates examples of \glspl{ibo} for a selection of materials.
These visualizations are qualitatively consistent with previously reported generalized Wannier orbitals derived from Pipek-Mezey type localization functionals.~\cite{Jonsson2017,Clement2021,Schreder2024,zhu2024wannier}

The \glspl{iao} can be constructed from any set of atomic functions $| f_\mu \rangle$ via the projection:
\begin{equation}
|\mu^{\text{IAO}}\rangle = (  \mathds 1 + \mathcal O - \widetilde{\mathcal O} ) | f_\mu \rangle \;,\label{eq:IAOs}
\end{equation}
where $\mathcal O$ is the projector onto the occupied space and $\widetilde{\mathcal O}$ projects onto the space spanned by occupied orbitals from a minimal atomic basis. These projectors are defined as:
\begin{equation}
\mathcal O = \sum_i^{N_{\text{occ}}} | \chi_i \rangle \langle \chi_i | \;, \quad \widetilde{\mathcal O} = \sum_i^{N_{\text{occ}}} | \widetilde\chi_i \rangle \langle \widetilde\chi_i | \;,
\end{equation}
with the orbitals $|\widetilde{\chi}_i\rangle$ given by
\begin{equation}
| \widetilde\chi_i \rangle = \text{orth}\left[\, \sum_{\mu\nu} | f_\mu \rangle \, S^{-1}_{\mu\nu}  \,  \langle f_\nu | \chi_i \rangle \, \right]\;,
\end{equation}
where $S_{\mu\nu} = \langle f_\mu | f_\nu \rangle$ is the overlap matrix of the atomic functions and "orth" denotes orthogonalization.
For the atomic functions $| f_\mu \rangle$ we use \gls{dft} orbitals of the free atoms.~\cite{Schafer2021}
While our definition of Intrinsic Atomic Orbitals (IAOs) in Eq. (\ref{eq:IAOs}) differs from Knizia's original formulation, they are equivalent when the minimal atomic basis is a subspace of the main basis. This condition is satisfied for a plane wave basis as a main basis, as the minimal atomic basis is also represented within it.
The minimal atomic basis orbitals $|\widetilde\chi_i\rangle$ approximate the occupied orbitals, while the exact occupied mean-field orbitals $|\chi_i\rangle$ are obtained from a preceding mean-field calculation in the plane wave basis.
The term $\mathcal O - \widetilde{\mathcal O}$ in Eq.~(\ref{eq:IAOs}) augments the atomic functions to form the \glspl{iao}, ensuring completeness of the occupied space.
As long as no occupied orbital is orthogonal to the atomic functions, i.e., $\sum_\nu \langle f_\nu | \chi_i \rangle \neq 0$, $\forall i$, the \glspl{iao} form an exact atom-centered basis for the occupied space. 
%Moreover, if the density spillage of the approximate orbitals $|\widetilde\chi_i\rangle$ is small, the \glspl{iao} retain a shape close to the original atomic functions $|f_\mu\rangle$, making them a localized and exact minimal basis for the occupied space.

%%%%%%%%%%%%%%%%%%%%%%%%%%%%%%%%%%%%%%%%%%%%%%%%%%%%%%
%%%%%%%%%%%%%%%%%%%%%%%%%%%%%%%%%%%%%%%%%%%%%%%%%%%%%%
%      Construction
%%%%%%%%%%%%%%%%%%%%%%%%%%%%%%%%%%%%%%%%%%%%%%%%%%%%%%
%%%%%%%%%%%%%%%%%%%%%%%%%%%%%%%%%%%%%%%%%%%%%%%%%%%%%%
\subsection{Riemannian Construction of Intrinsic Bond Orbitals} \label{sec:construct}

Efficient optimization algorithms are vital for the success of localization methods, relying on the optimization of an orbital-dependent cost function $\mathcal{L}$.
As described previously in Sec. \ref{sec:theory}, we employ the \gls{ibo} method. Optimization is performed using a Riemannian geometry approach, exploiting the topological properties of the unitary group to preserve the unitary constraint inherently. The search directions are translated to geodesics on the manifold, leading to more efficient optimization steps. Early works on these topics were conducted, for example, by Luenberger and Gabay~\cite{luenberger1972,gabay1982minimizing}.

\subsubsection{Riemannian optimization under unitary constraint}\label{sec:riemann_optimization}

We opted for a Riemannian optimization approach due to its inherent suitability for handling unitary matrix constraints. 
Unlike traditional Euclidean methods that struggle to maintain unitarity and often suffer from slow convergence, Riemannian optimization operates directly on the manifold of unitary matrices. 
An illustrative comparison of how Riemannian and Euclidean algorithms operate under the unitary constraint were provided by Abrudan et al. in Ref. \citenum{Abrudan2008}.
The Riemannian approach respects the inherent ``curved space'' nature of the parameter space, allowing optimization along geodesics---the most efficient paths on this manifold. 
Furthermore, by recognizing that unitary matrices form a Lie group under multiplication, we leverage the algebraic properties of this group to ensure unitarity is preserved throughout the optimization process.
This avoids the need for costly restoration steps or penalty functions, leading to more accurate and efficient convergence.

Riemannian optimization leverages the theory of optimization and concepts of differential geometry, more specifically Riemannian manifolds.
We follow the works of Abrudan et al.~\cite{Abrudan2008,Abrudan2009} and Huang et al.~\cite{huang2015broyden,huang2013optimization}.
Another key work to mention in this context is the study of Edelman et al. in Ref. \citenum{Edelman2006}.

\subsubsection{Unconstrained Optimization} \label{sec:optimization}
In this section, we introduce the concept of unconstrained line search algorithms, which are later adapted for application on manifolds. A minimum (or maximum) of some function $\mathcal L(\bm U)$ is approached iteratively, where $\bm U$ represents an abstract vector in the parameter space. 
The optimization algorithms we compare are \acrfull{cg}, \acrfull{lbfgs} and \acrfull{sa} solvers, in this paper we focus particularly on the first two, as \gls{sa} has proven to be clearly inferior in our calculations and in Refs.~\citenum{Clement2021,zhu2024wannier}.  Line search algorithms select a suitable direction in parameter space in a first step and subsequently determine an optimal step size along that chosen path. A detailed treatment of these topics can be found in Ref.~\citenum{nocedal1999numerical}.
Without constraints, these algorithms are unrestricted within the respective parameter space, and follow the general update formula
\begin{equation}
    \bm U_{k+1} = \bm U_k + \alpha_k \bm H_k,
\end{equation}
where the iterates $\bm U_{k+1}$ and $\bm U_k$ are estimates of the desired extremum, $\alpha_k$ is the step size $\bm H_k$ is the search direction. 

For \gls{sa}, the search direction $\bm H$ is chosen as the gradient $\bm\nabla \mathcal L(\bm U_k)$. The \gls{cg} search direction is calculated according to the formula:
\begin{equation}
    \bm H_{k} = \bm\nabla \mathcal L(\bm U_k) + \beta_k \bm H_{k-1},
\end{equation}
where $\beta$ is a weighting factor that uses information from the previous step.
Based on the work from Lehtola et al.~\cite{lehtola2013unitary} we use the \gls{pr} formula for the factor $\beta_k$~\cite{polak1969pr, polak1971computational}.
The initial search direction is $H_{0} = \bm\nabla \mathcal L(\bm U_0)$. 
\gls{bfgs}\cite{broyden1970convergence,fletcher1970new,goldfarb1970family,shanno1970conditioning} is a quasi-Newton algorithm that mimics Newton's method of minimizing the second-order Taylor series of the cost function. The Newton search direction is $\bm H_k=-\bm B_k \bm\nabla \mathcal L(\bm U_k)$ with the inverse Hessian $\bm B_k$. For high-dimensional problems, the computational cost of calculating the Hessian or its inverse is usually prohibitively high, so quasi-Newton algorithms aim for an accurate approximation. 
The \gls{lbfgs} approximation is given by:
\begin{equation} \label{eq:BFGS_formula}
    \bm B_{k+1} = (\bm I - \rho_k \bm y_k \bm s_k^T) \bm B_k (\bm I - \rho_k \bm y_k \bm s_k^T) + \rho_k \bm s_k \bm s_k^T
\end{equation}
where $\bm I$ is the identity,
\begin{equation}
    \rho_k = \frac{1}{\bm y_k^T \bm s_k}, 
\end{equation}
and
\begin{equation} \label{eq:s_and_y}
    \bm s_k = \bm U_{k+1}- \bm U_k, \quad \bm y_k = \bm\nabla \mathcal L_{k+1} - \bm\nabla \mathcal L_k
\end{equation}

A requirement for the existence of a solution is the so-called curvature condition $\bm s_k^T \bm y_k > 0$, which ensures that the Hessian is positive definite and therefore invertible. This can be ensured by using a step size algorithm that is based on the Wolfe conditions~\cite{nocedal1999numerical}, for example.
Note that the vectors in equations \ref{eq:BFGS_formula} to \ref{eq:s_and_y} are not necessarily one-dimensional objects. As discussed later, in our use case we treat unitary matrices as abstract vectors with the corresponding Frobenius product serving as inner product.

\gls{lbfgs}~\cite{liu1989lbfgs} is an approximation of \gls{bfgs}, designed specifically for high-dimensional problems. While \gls{bfgs} stores $\bm B$ explicitly, the limited memory version \gls{lbfgs}~\cite{liu1989lbfgs} approximates equation (\ref{eq:BFGS_formula}) iteratively, therefore only retaining vectors $\bm y$ and $\bm s$ from a fixed number of previous iterations (memory). In our case, $\bm y$ and $\bm s$ are matrices of size $n \times n$, making $\bm B$ of size $n^4$. This large scaling restricts \gls{bfgs} to small systems, while \gls{lbfgs} is usually the method of choice for large-scale problems with a high-dimensional parameter space. With increasing memory size, the \gls{lbfgs} approximation approaches \gls{bfgs} and if every step is stored they are mathematically equivalent.

When setting the memory size to 1, \gls{lbfgs} is closely related to \gls{cg} methods, which also memorize the gradient at the previous point to update the search direction~\cite{nocedal1999numerical}.

After finding a search direction applying one of the above methods, a suitable step size has to be selected to determine an exact point along that path. Step size algorithms are in general independent of the way search directions are selected, although some are more suitable than others. Abrudan et al. suggest interpolating along the search path and calculating the maximum (minimum) of the resulting polynomial~\cite{Abrudan2009} to obtain a reasonable estimate.

\subsubsection{Riemannian Geometry} \label{sec:riemann_geometry}
A very elegant way to impose constraints on parameters is to exploit topological properties of the parameters. For a more detailed treatment of the concepts in this chapter, especially in the context of optimization, consider the references~\cite{qi2011numerical,huang2013optimization, Abrudan2009}.

We introduce Riemannian manifolds, smooth manifolds equipped with a metric.
In general, a smooth manifold $\mathcal{M}$ is a topological space that fulfills special requirements regarding distance, neighborhood and differentiability. 
To each point $\bm U \in \mathcal{M}$, a tangent space $T_{\bm U}\mathcal{M}$ is attached, i.e., the set of all possible tangent vectors at that point. 

Consider a smooth curve 
\begin{equation} \label{eq:curve_on_manifold}
    \gamma(t):\mathbb{R}\rightarrow \mathcal{M}, \qquad \gamma(0) = \bm U. 
\end{equation} 
If $\mathcal{M}$ is a submanifold of Euclidean space, a tangent vector $\bm X$ to $\mathcal{M}$ at point $\bm U$ is intuitively defined as the derivative of this curve at $t=0$, 
\begin{equation} \label{eq:tangent_vector}
\bm X_{\bm U} \equiv \frac{d}{dt}\gamma(t) |_{t=0}  
\end{equation}
In this sense, a tangent vector defines the direction of a curve on the manifold.

A textbook example for a mapping procedure between tangent spaces and the manifold itself is the exponential map (see Eq.(\ref{eq:geodesics})), which enables movement along curves.

Riemannian manifolds are equipped with a Riemannian metric, defined on each tangent space as inner product $g(\bm X,\bm Y) = \langle \bm X,\bm Y \rangle$, where $\bm X,\bm Y$ are tangent vectors.

\subsubsection{The unitary group $U(n)$}
A key property of unitary $n \times n$ matrices is that they form a Lie group $U(n)$, with matrix multiplication as a group action. The tangent space of the point at unity is highlighted as the Lie algebra of the group, $T_{\bm I}U(n) \equiv \mathfrak{u}(n)$, consisting of all skew-hermitian $n \times n$ matrices. 

The group action defines two maps, known as \textit{right translation} and \textit{left translation}, meaning the multiplication of a point $\bm V \in U(n)$ by another point on the right:

\begin{equation}
\begin{matrix}
     R_U: & U(n) & \rightarrow & U(n) \\
     & \bm V & \mapsto & \bm V \bm U \equiv & \bm V'
\end{matrix}
\end{equation}
and equivalently for left translation.

A tangent vector $X_{\bm V} \in T_{\bm V}U(n)$ can be translated in the same way to another tangent space $T_{\bm V \bm U}U(n)$:
\begin{equation} \label{eq:tangent_translation}
\begin{matrix}
    R_{U*}: & T_{\bm V}U(n) & \rightarrow & T_{\bm V'}U(n)\\
    & \bm X_{\bm V} & \mapsto & \bm X_{\bm V} \bm U & \equiv & \bm X_{\bm V'}
\end{matrix}
\end{equation}

Importantly, these translations are isometries with respect to the Riemannian metric, so distances are preserved, allowing for the simple movement of curves and tangent vectors between points on the manifold. Following equation \ref{eq:tangent_translation}, every vector in the Lie algebra $X_{I} \in \mathfrak u$ can be moved to any tangent space $T_{\bm U}U(n)$ by multiplication with $\bm U$ from the right, $\bm X_{\bm U} = \bm X_{\bm I}\bm U$, and vice versa every tangent vector can be easily translated to the Lie algebra: $\bm X_{\bm I} = \bm X_{\bm U} \bm U^\dagger$. This makes the Lie algebra a very convenient choice for calculations involving multiple tangent vectors.

The exponential mapping $\exp:\mathfrak{u} \rightarrow U(n)$ maps an element $\bm X \in \mathfrak{u}$ to the group, given by the matrix exponential 
\begin{equation} \label{eq:geodesics}
    \text{exp}(\alpha\bm X)=\bm\gamma(\alpha),
\end{equation} 
where the curve $\bm\gamma:\mathbb{R}\rightarrow U(n)$ is a parameterized geodesic, the shortest path between two points of the group. This can be understood as taking a direction $\bm X$ and moving along the corresponding geodesic curve.

The concepts in this chapter are also valid for the orthogonal group $O(n)$, consisting of orthogonal matrices as elements, while skew-symmetric matrices form the Lie algebra.

\subsubsection{Optimization on the Unitary Group}
\label{sec:optimization_on_group}
Combining the previously discussed ideas, optimization algorithms originally designed as unconstrained in the Euclidean parameter space can be generalized to Riemannian manifolds. Equipped with the Frobenius inner product as Riemannian metric, $g(\bm X, \bm Y) = Tr(\bm X^\dagger \bm Y)$, the unitary group $U(n)$ forms a Riemannian manifold.

According to Abrudan et al.~\cite{Abrudan2008}, the gradient of a function $\mathcal L : U(n) \rightarrow \mathbb{R}$ at some point $\bm U \in U(n)$ is given by
\begin{equation}
    \bm\nabla \mathcal L(\bm U) = \bm\Gamma - \bm U \bm\Gamma^\dagger \bm U
\end{equation}
where $\bm\Gamma \equiv d\mathcal L / du_{ij}$.
Subsequently, this gradient is translated to the Lie algebra via right translation:
\begin{equation} \label{eq:riemannian_gradient}
    \bm G(\bm U) \equiv \bm\nabla\mathcal L(\bm U) \bm U^\dagger = \bm\Gamma \bm U^\dagger - \bm U \bm\Gamma^\dagger
\end{equation}
The algorithms \gls{sa}, \gls{cg} and \gls{bfgs} are now introduced following section \ref{sec:optimization}, utilizing the translated gradient $\bm G(\bm U)$ to obtain the search direction $\bm H_k$. This vector is mapped to the group using the exponential map in equation (\ref{eq:geodesics}), the emanating curve $\text{exp}(\alpha\bm H_k)$ is transported to $\bm U_k$ to obtain $\bm U_{k+1}$:
\begin{equation}
    \bm U_{k+1} = \text{exp}(\alpha\bm H)\bm U_k 
\end{equation}
where the scaling factor $\alpha$ serves as step size.

In the case of \gls{lbfgs}, the fact that $U(n)$ is a Lie group is especially advantageous. Vectors $y_i$ and $s_i$ do not need to be transported to the new iterate to calculate the search direction, as calculations can be performed in the Lie algebra.

\section{Computational Methods \& Implementation}  \label{sec:implement}

\begin{table}
\centering
\caption{Used \texttt{POTCAR} files, defining the \gls{paw} pseudopotential as well as the plane wave cutoff. For every material the largest \texttt{ENMAX} value was scaled by the factor $1.25$ to define the plane wave cutoff \texttt{ENCUT}.}
\begin{tabular}{llll}
\hline\hline
Element & \texttt{POTCAR} header           & valence & \texttt{ENMAX} (eV) \\
\hline
H & \texttt{PAW\_PBE H\_GW 21Apr2008}     & $1 {\text s}^1$ & 300.000 \\
Li & \texttt{PAW\_PBE Li\_AE\_GW 25Mar2010} & $1 {\text s}^2 2 {\text p}^1$ & 433.699 \\
B & \texttt{PAW\_PBE B\_GW\_new 26Mar2016} & $2 {\text s}^2 2 {\text p}^1$ & 318.614 \\
C & \texttt{PAW\_PBE C\_GW\_new 19Mar2012} & $2 {\text s}^2 2 {\text p}^2$ & 413.992 \\
N & \texttt{PAW\_PBE N\_GW\_new 19Mar2012} & $2 {\text s}^2 2 {\text p}^3$ & 452.633 \\
O & \texttt{PAW\_PBE O\_GW\_new 19Mar2012} & $2 {\text s}^2 2 {\text p}^4$ & 434.431 \\
F & \texttt{PAW\_PBE F\_GW\_new 19Mar2012} & $2 {\text s}^2 2 {\text p}^5$ & 480.281 \\
Na & \texttt{PAW\_PBE Na\_sv\_GW 11May2015} & $2 {\text s}^2 2 {\text p}^6 3 {\text p}^1$ & 372.853 \\
Mg & \texttt{PAW\_PBE Mg\_GW 13Apr2007}    & $3 {\text s}^2$ & 126.143 \\
Al & \texttt{PAW\_PBE Al\_GW 19Mar2012}    & $3 {\text s}^2 3 {\text p}^1$ & 240.300 \\
Si & \texttt{PAW\_PBE Si\_GW\_nc 03Jul2013} & $3 {\text s}^2 3 {\text p}^2$ & 319.379 \\
P & \texttt{PAW\_PBE P\_GW 19Mar2012}     & $3 {\text s}^2 3 {\text p}^3$ & 255.040 \\
Cl & \texttt{PAW\_PBE Cl\_GW 19Mar2012}    & $3 {\text s}^2 3 {\text p}^5$ & 262.472 \\
Ti & \texttt{PAW\_PBE Ti\_sv\_GW 05Dec2013} & $3 {\text s}^2 3 {\text p}^6 3 {\text d}^4$ & 383.774 \\
Ga & \texttt{PAW\_PBE Ga\_GW 22Mar2012}    & $4 {\text s}^2 4 {\text p}^1$ & 134.678 \\
Ge & \texttt{PAW\_PBE Ge\_GW 04Okt2005}    & $4 {\text s}^2 4 {\text p}^2$ & 173.807 \\
As & \texttt{PAW\_PBE As\_GW 20Mar2012}    & $4 {\text s}^2 4 {\text p}^3$ & 208.702 \\
\hline\hline
\end{tabular}
\label{tab:potcar}
\end{table}

The \gls{cg} and \gls{sa} solvers are implemented in the publicly available Julia package Lucon.jl (Loss optimization under unitary constraint)~\cite{lucon} and in \gls{vasp} as reported in Ref. ~\citenum{Schafer2021b}.
All considerations with the \gls{cg} solver are performed using the Polak-Ribière update factor.~\cite{polak1969pr}
The implementation of the \gls{lbfgs} solver was included in a development version of \gls{vasp} in the scope of this work.
Pseudocode for this algorithm is shown in Alg. \ref{alg:LBFGS}, following the work of Huang et al. and Nocedal et al.~\cite{huang2015broyden,nocedal1999numerical}.
The two-loop recursion was developed by Nocedal et al.~\cite{nocedal1999numerical} and efficiently computes the \gls{lbfgs} search direction. For step size calculations, we utilize the method developed by Abrudan et al.~\cite{Abrudan2009}.
To validate our implementation and confirm our results, we repeated all calculations using the manopt.jl package by Bergmann et al.~\cite{bergmann2022manopt,AxenBaranBergmannRzecki2023manifolds}, where we selected a step size algorithm based on the Hager-Zhang scheme~\cite{hagerzhang2005,Baran2024}.
The Euclidean derivative for the \glspl{ibo} reads
\begin{align}
&\Gamma^\text{PM}_{ij} = \frac{\partial \mathcal L^\text{PM}}{ \partial u^*_{ij}} \label{eq:GammaIJ} \\
&= 2\sum_A^{N_\text{atoms}} \sum_{k}^{N_{\text{occ}}} \langle \chi_i | \bm P_A | \chi_k \rangle   \, u_{kj}  \bigg| \sum_{lm}^{N_{\text{occ}}} \langle \chi_l | \bm P_A | \chi_m \rangle   \, u^*_{lj} \, u_{mj
} \bigg| \nonumber\;.
\end{align}

The \gls{hf} orbitals $| \chi_i \rangle$ are obtained from the plane-wave based \acrfull{vasp} \cite{Kresse1993,Kresse1996a,Kresse1996b} using the \gls{paw} method~\cite{Blochl1994}.
The \gls{paw} pseudopotentials use a frozen core and are provided as \texttt{POTCAR} files with \gls{vasp}, see Tab. \ref{tab:potcar}.
For each material, the largest \texttt{ENMAX} value was multiplied by a factor of 1.25, and then rounded up to the nearest multiple of ten to determine the plane wave cutoff \texttt{ENCUT} in units of eV. 
By scaling the default value (\texttt{ENMAX}) in this way, we ensure that we use a sufficiently large base for each material.
For example, the calculations for \ce{SiC} were performed using $\texttt{ENCUT} = \lceil \text{max}(413.992, 319.379)\cdot 1.25 \rceil_{10} = \lceil 517.490 \rceil_{10} = 520$, where $\lceil x \rceil_{10}$ denotes rounding $x$ up to the nearest multiple of 10.
The singularity of the Coulomb potential in reciprocal space is treated via the truncation method introduced by Spencer and Alavi in Ref. \citenum{Spencer2008}.
Supercells are considered using a $\bm \Gamma$-only sampling of the \gls{bz}.
The atomic structures for caffeine, benzene, coronene, graphene with flower defect, and silicon with interstitial defect can be found in the supplementary information~\cite{si}.

\begin{algorithm}
\caption{L-BFGS algorithm implementation} \label{alg:LBFGS}
\begin{algorithmic}

\State choose memory size m, break condition $\epsilon$
\State choose starting point $U_0$
\State calculate the initial gradient $G(U_0)$ from equation (\ref{eq:riemannian_gradient})
\State $s_i, y_i, \rho_i \gets 0$
\State $\lambda \gets 1$
\State $k \gets 0$
\Repeat
\State $k \gets k+1$
\Procedure{two-loop recursion}{see Ref. \citenum{nocedal1999numerical} for details}
\State $q \gets G(U_k)$
\For{$i=m,m-1,...,1$}
\State $a_i \gets \rho_i g(s_i,q)$
\State $q \gets q-a_i y_i $
\EndFor
%\State $\lambda \gets g(s_m,y_m)/g(y_m,y_m)$
\State $r \gets \lambda q$
\For{$i=1,...,m$}
\State $b \gets \rho_i g(y_i,r)$
\State $r \gets r + s_i (a_i - b)$
\EndFor
\State $H_k \gets -r$
\EndProcedure
\State perform step size algorithm to obtain $\alpha_k$
\State $U_{k+1} = \text{exp}(\alpha_k H_k) U_k$
\State calculate $G(U_{k+1})$
\For{$i=1,...,m-1$}
\State $s_i \gets s_{i+1}, \quad$
$y_i \gets y_{i+1}, \quad$
$\rho_i \gets \rho_{i+1}$,
\EndFor
\State $s_m \gets \alpha_k H_k$
\State $y_m \gets G(U_{k+1})-G(U_k)$
\State $\rho_m \gets 1/g(s_m,y_m)$
\State $\lambda \gets g(s_m,y_m)/g(y_m,y_m)$
\Until{$||G(U_k)|| < \epsilon$}
\end{algorithmic}
\end{algorithm}

We also implemented the \gls{diis} technique to investigate its potential for accelerating convergence to the optimum. 
The \gls{diis} technique is a mixer that seeks to find optimal linear combinations of previous iteration steps. 
This technique is well-established for accelerating iterative solvers in finding the \gls{hf} ground state.
When finding an optimal unitary matrix, this matrix must be parametrized to construct linear combinations of previous solutions, resulting in a new unitary matrix. 
While several parametrizations exist~\cite{Shepard2015}, we used the exponential parametrization, which was already successfully applied for the rotation of orbitals in previous works~\cite{Ionova1995}.
In the case of unitary (orthogonal) rotations, we write $U=\EuE^{\ImI \Theta}$ ($U=\EuE^{\Theta}$) with the hermitian (skew-symmetric) matrix $\Theta$ containing the rotation parameters.
Note, that this consideration no longer follows the idea of a Riemannian optimization, but is necessary to mix parameters (here $\Theta$) in the \gls{diis} mixer.
%Assuming orthogonal rotations $U$ we calculate their parametrization as $\Theta = -\ImI \log(U)$, where the parameter matrix $\Theta$ is hermitian.
Our implementation was modeled after the documentation by C. D. Sherrill~\cite{sherrill2004some}. 
Accordingly, we define the error vectors of the \gls{diis} scheme as $\Delta_i = \Theta_i - \Theta_{i-1}$ and find the optimal parameters $\Theta_\text{opt} = \sum_{i=1}^n \tau_i \Theta_i$ by minimizing the Frobenius norm of $\Delta = \sum_{i=1}^n \tau_i \Delta_i $, where $n$ represents a fixed history size. 
The optimal parameters $\Theta_\text{opt}$ themselves are never added to the history to avoid linear dependencies.

\section{Results} \label{sec:results2}

We performed computations for several molecules, molecular crystals, bulk solids and systems with broken translational symmetry. A special focus was on large supercells.
If not stated otherwise, the calculations were initialized with random unitary matrices and a break condition for the gradient norm of $||\bm G|| = \sqrt{\langle \bm G, \bm G\rangle} < 10^{-5}$ was chosen as the convergence criterion. 
We measure the \textit{performance} of an algorithm by the number of iterations required to reach convergence.

\begin{figure}
    \centering
    \includegraphics{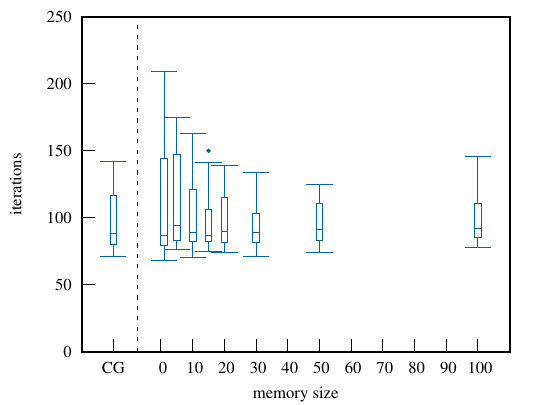}
    \caption{Box-and-whisker plot of the number of required iterations against the number of memorized \gls{lbfgs} steps compared to \gls{cg} for a graphene supercell (162 atoms) with flower defects.
    }
    \label{fig:it_against_memory}
\end{figure}

Surprisingly, specifically for periodic systems, large \gls{lbfgs} memory sizes do not necessarily lead to improved performance for \gls{ibo} localization, as shown in Fig.\ref{fig:it_against_memory} for a graphene flower defect system (supercell with 324 occupied orbitals). 
However, the statistical variance of the required number of iterations decreases with higher memory, while the increase in computational cost is negligible. 
If not stated otherwise, a fixed memory size of 20 is used for our \gls{lbfgs} calculations. 
Note that the performance of \gls{cg} and \gls{lbfgs} is similar for this example, an observation that is consistent across all periodic systems tested. 

\begin{figure}
    \centering
    \includegraphics{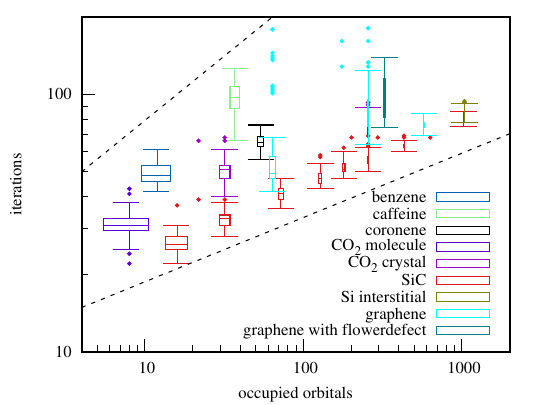}
    \caption{Box-and-whisker plot of the \gls{lbfgs} iterations against occupied orbitals $n$ for several systems and supercell sizes. The dashed lines are proportional to $\sqrt{n}$ and $\sqrt[4]{n}$ respectively, giving an idea of scaling.
    }
    \label{fig:it_against_system_size}
\end{figure}

Fig. \ref{fig:it_against_system_size} shows the median number of necessary iterations against the system size for a selected set of systems using the \gls{lbfgs} solver.
The scaling of the iterations with system size is roughly proportional to the fourth root of the number of occupied orbitals $n$, i.e. sublinear, illustrated by the dashed lines.

\begin{table}
    \centering
    \begin{tabular}{lrrrr}
        \hline\hline
        material / molecule & \#occ    & \gls{lbfgs} & \gls{cg} & \gls{sa} \\ \hline
        benzene     & 12      & 49      & 83      & 7093    \\ %\hline
        caffeine     & 37      & 97      & 132     & 3217    \\ %\hline
        coronene     & 54      & 65      & 85      & 671     \\ %\hline
        \ce{CO2} molecule      & 8       & 31      & 38      & 171     \\ %\hline
        \ce{CO2} crystal     & 32      & 51      & 53      & 269     \\ 
            & 256     & 73      & 81      & 316     \\ %\hline
        \ce{SiC}      & 16      & 26      & 26      & 54      \\
             & 32      & 33      & 33      & 68      \\
              & 72      & 41      & 41      & 87      \\
          & 128     & 47      & 45      & 102     \\
         & 180     & 51      & 50      & 110     \\
         & 256     & 56      & 53      & 120     \\
         & 432     & 63      & 58      & 134     \\ %\hline
        % & 1024    & 79      & 70      &         \\ \hline
        defect \ce{Si}    & 1040    & 84      & 74      & 154     \\ %\hline
        graphene    & 64      & 50      & 51      & 124     \\
            & 256     & 89      & 78      & 197     \\
            & 576     & 76      & 67      & 155     \\ %\hline
        flower defect graphene     & 324     & 90      & 88      & 265     \\  \hline\hline
    \end{tabular}
    \caption{Median number of required iterations for \gls{lbfgs}, \gls{cg} and \gls{sa} solvers for several systems of various cell size (number of occupied orbitals).}
    \label{tab:it_against_occ}
\end{table}

In Tab. \ref{tab:it_against_occ}, the median number of required iterations are listed for \gls{lbfgs}, \gls{cg} and \gls{sa} algorithms. 
Interestingly, for supercells with broken symmetry, \gls{lbfgs} and \gls{cg} show a performance similar to that for the pristine case, as the results indicate for the flower defect graphene and \ce{Si} with interstitial defects. 
\gls{lbfgs} and \gls{cg} outperform \gls{sa} for all test systems, \gls{lbfgs} has an advantage over \gls{cg} only for molecules. 

\begin{table*}
    \centering
    \begin{tabular}{l|r|r|r|r|r}
        \hline\hline
        material & \#occ      & \multicolumn{2}{l|}{largest maximum} & \multicolumn{2}{l}{second largest maximum}      \\
                 &       & iterations    & cost per \#occ         & iterations    & cost per \#occ  \\
        \hline
        graphene & 64  & 49 (92\%) & 1.1357 & 112 (8\%)    & 1.1278 \\
         & 256 & 87 (90\%) & 1.1346    & 126 (7\%)    & 1.1293 \\
         & 576 & 76 (100\%) & 1.1430  & 0 (0\%)             & \\
        flower defect & 324  & 83 (62\%) & 1.1350   & 119 (38\%)   & 1.1341 \\
        \hline\hline
    \end{tabular}
    \caption{For several graphene supercells, the algorithms not always converge to the same value of the cost function. The percentage of runs (out of 60 each) converging to the respective maximum is given in brackets. Notably, when converging to a lower value, the number of iterations is considerably higher. These outliers are also visible in Fig. \ref{fig:it_against_system_size} for the graphene cells. }
    \label{tab:cost_comparison}
\end{table*}  

Notably, some graphene cells exhibit outliers with a substantially higher number of iterations, approaching other local extrema. This behavior, visible in Fig. \ref{fig:it_against_system_size}, is further quantified in Tab. \ref{tab:cost_comparison} for the \gls{lbfgs} algorithm. For example, flower defect graphene calculations (324 occupied orbitals) converge to a slightly worse maximum in 38\% of runs. Similar observations were made with the manopt.jl package considering \gls{cg} and \gls{lbfgs}, using a different line search method. 

\begin{table}
    \centering
    \begin{tabular}{llrrrr}
        \hline\hline
        material & oxide & \#occ & \gls{ibo} & FB & VN \\
        \hline
        \ce{SiC}  &  non-oxide               & 256 & 53     & 140    & 39  \\
        \ce{CO2}  &  molecular oxide crystal & 256 & 81     & 546    & 294    \\
        \ce{SiO2} ($\alpha$-quartz) &  non-metal oxide	     & 288 & 155    & 496    & 154    \\
        \ce{TiO2} (rutile) &  metal oxide             & 288 & 678    & 1628    & 605 \\
        \ce{MgO}  &  metal oxide             & 288 & 1737   & 1464    & 365    \\
        \hline\hline
    \end{tabular}
    \caption{Comparison of the the average number of necessary iterations for metal oxides and a set of other materials employing the \acrfull{ibo}, \acrfull{fb}, and \acrfull{vn} localization functionals. Supercells containing a comparable number of occupied orbitals were considered. The average was calculated from 4 runs initialized with random unitary matrices.}
    \label{tab:iteroxide}
\end{table}

\begin{figure}
    \centering
    \includegraphics{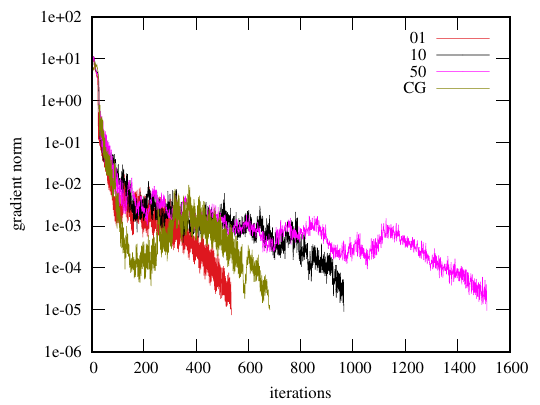}
    \caption{Convergence of \gls{lbfgs} with several memory size settings and \gls{cg}, for a \ce{TiO2} supercell (72 atoms, 288 occupied orbitals), the unity matrix serves as starting point (i.e. starting from bloch orbitals).
    }
    \label{fig:grad_against_it_TiO2_72}
\end{figure}

As listed in Tab. \ref{tab:iteroxide}, metal oxides require about one order of magnitude more iterations for a fixed convergence threshold than other systems. 
Here, all considered localization functionals \gls{ibo}, \gls{fb} and \gls{vn} show a similar trend. 
The convergence behavior of the \gls{lbfgs} optimization is illustrated in Figure \ref{fig:grad_against_it_TiO2_72}, which displays the gradient norm per iteration for a \ce{TiO2} supercell containing 72 atoms and 288 occupied orbitals. An initial rapid reduction in the gradient norm is observed within the first 50-100 iterations, transitioning to a slower, more irregular decrease accompanied by significant oscillations.

\begin{figure}
    \centering
    \includegraphics{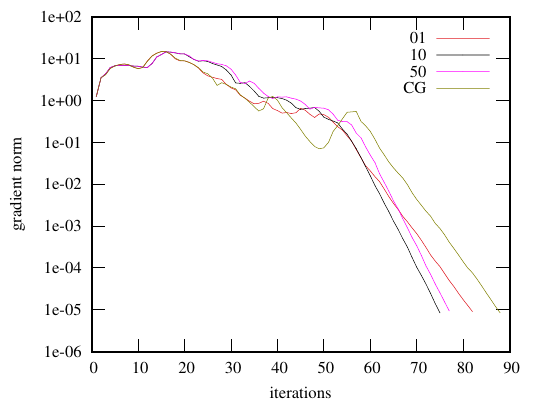}
    \caption{Convergence of \gls{lbfgs} with several memory size settings and \gls{cg}, for a graphene supercell (162 atoms) with flower defects, the unity matrix serves as starting point (i.e. starting from bloch orbitals). 
    }
    \label{fig:grad_against_it_flower defect_162}
\end{figure}

In contrast, a typical convergence pattern of non-oxides is shown in Fig. \ref{fig:grad_against_it_flower defect_162} using the aforementioned graphene flower defect supercell as an example. From a certain iteration step onwards (in this case somewhere between 50 and 60 iterations), convergence is consistently exponential.

\begin{figure}
    \centering
    \includegraphics{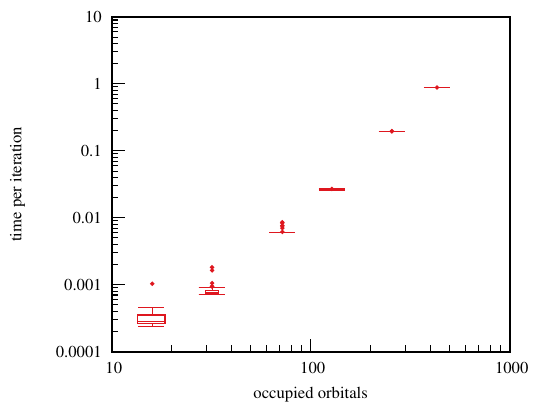}
    \caption{\gls{lbfgs} runtime per iteration in seconds against occupied orbitals, $n$, of \ce{SiC} supercells, showing a $n^3$ scaling.
    }
    \label{fig:time_against_occ}
\end{figure}

The largely cubic scaling of the \gls{lbfgs} runtime per iteration and the system size is depicted in Fig. \ref{fig:time_against_occ} for \ce{SiC}. Except for very small cells, where the impact of several inexpensive routines is visible, the runtime scales proportionally to $n^3$, where $n$ is the number of occupied orbitals. This behavior is expected, as $n$ determines the size of most of the involved matrices and consequently the cost of matrix operations.
\gls{cg} and \gls{sa} are only marginally faster, on average by $3.5\%$ and $4.1\%$, respectively, disregarding the two smallest cells. It can be stated that the additional complexity of \gls{lbfgs} is insignificant in relation to the cost of other routines like gradient calculation or line search algorithm.

\begin{figure}
    \centering
    \includegraphics[width=0.6\linewidth]{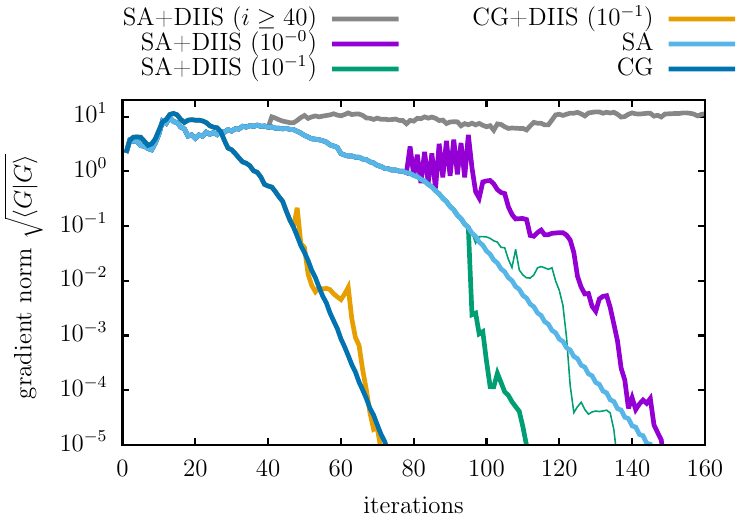}
    \caption{Convergence of the DIIS solver (memory size 10) for a \ce{SiC} supercell with 256 occupied orbitals. The gradient norm or iteration threshold for initiating DIIS is shown in brackets. Thin lines represent the case when the DIIS starts with an empty history. All calculations use the identity as the starting point, i.e. Bloch orbitals.}
    \label{fig:diis}
\end{figure}

In order to assess wether the \gls{diis} technique can accelerate the convergence of our Riemannian solvers, we considered a supercell of \ce{SiC} containing 256 occupied orbitals.
We start the \gls{diis} mixer when the gradient norm fell below a certain threshold using a fixed history size of 10.
As evident in Fig. \ref{fig:diis} the \gls{diis} technique can beat the convergence of the \gls{sa} algorithm using a threshold of $10^{-1}$.
A clear distinction is visible between an already filled history (thick lines) and the case when the mixer starts from an empty history (thin lines).
Increasing the threshold to $10^{0}$ leads to a slightly worse convergence behavior.
When the \gls{diis} technique is activated following a certain number of iterations (here $i\geq 40$), it exhibits poor convergence behavior from a suboptimal initial state, necessitating 719 iterations to achieve convergence.
Furthermore, the \gls{diis} mixer is unable to accelerate the convergence of the \gls{cg} solver in combination with the \acrfull{pr} update factor.
This also applies, unfortunately, to the challenging case of metal oxides.
We note that the \gls{diis} solution is always updated by the \gls{sa} solver, i.e. the label ``CG+DIIS''  in Fig. \ref{fig:diis} denotes a \gls{cg} solution until the threshold is reached, followed by the \gls{sa} solver with \gls{diis} mixing.
This is due to the fact that updating the \gls{diis} solution using \gls{cg} with \gls{pr} factors consistently leads to a non-converging behavior.

\section{Discussion} \label{sec:discussion2}

In this first part of the paper, we consider the performance of various solvers, avoiding any bias such as initial guesses.
We observed that both \gls{lbfgs} and \gls{cg} exhibited similar performance and significantly outperformed the \gls{sa} solver within their respective Riemannian formulations.
The comparable performance of our \gls{cg} and \gls{lbfgs} implementations, coupled with the latter's limited sensitivity to memory size, suggests potential limitations in the \gls{lbfgs} Hessian approximation specifically within the context of \gls{ibo} localization. 
This contrasts with \gls{fb} localization, where larger \gls{lbfgs} memory sizes have been shown to improve performance and \gls{lbfgs} consistently outperforms \gls{cg}.

\gls{lbfgs} iterations are only marginally slower than \gls{cg} and \gls{sa} in runtime measurements, dispelling a potential disadvantage and indicating the runtime dominance of other routines like gradient calculation or step size search. Runtime per iteration scales cubically with the number of occupied orbitals. 

For graphene supercells, both pristine and defect-containing, we observed that multiple stochastically initialized runs converged to distinct, suboptimal local maxima of the cost function. 
This problem is consistent across all solvers tested and indicates a significant presence of local extrema and saddle points in the optimization landscape, a common challenge for high-dimensional cost functions.

We also observe that the localization procedure for the metal oxides \ce{MgO} and \ce{TiO2} requires significantly more iterations to converge compared to other systems examined.
While these materials exhibit strong ionic character, this characteristic alone does not explain the observed slow convergence.
Specifically, we did not encounter similar convergence difficulties with other ionic systems such as \ce{LiF} and \ce{NaCl}, which share the same crystal structure as \ce{MgO}.
The primary distinguishing features of \ce{MgO} and \ce{TiO2} compared to the other systems are the ionic nature in combination with the -2 charge of the anion and the metallic nature of the cation.

Despite its effectiveness in accelerating the convergence of the \gls{sa} solver at sufficiently low gradient norms (below $10^{-1}$), the \gls{diis} mixer did not yield a comparable improvement for the \gls{cg} solver.
Furthermore, the convergence difficulties encountered with metal oxides remained unaffected by the application of the \gls{diis} mixer.

Our results also present a noteworthy divergence from those of Clement et al.~\cite{Clement2021}, as briefly noted in the Introduction.
They report a significant performance advantage for the Riemannian \gls{lbfgs} over the \gls{cg} solver for any of the materials considered in their work.
This discrepancy is particularly pronounced in graphene, where Clement et al. reported approximately $3\cdot 10^3$ iterations for \gls{cg} convergence of \gls{pm} orbitals, while our \gls{cg} implementation convergences in less than $10^2$ iterations.
Notably, our \gls{lbfgs} results align with those of Clement et al., showing comparable performance.

We speculate that the discrepancies for the \gls{cg} solver could be due to the following different technical features of the implementations. 
For instance, while our approach uses large supercells sampled at the $\bm \Gamma$-point only, Clement et al. employs small unit cells in combination with fine k-point sampling of the \gls{bz}.
The $\bm \Gamma$-point only approach allows us to use real-valued orbitals in the real-space basis, avoiding the gauge freedom with complex phases and the challenging search for a smooth gauge~\cite{Marzari2012}.
Furthermore, while we restrict the optimization to valence electrons using a frozen core in combination with \gls{paw} pseudopotentials in the plane wave basis, Clement et al. follows an all-electron approach using Gaussian basis sets. 
A significant performance gain when restricting the optimization to valence electrons was already reported by Zhu et al. in Ref. \citenum{zhu2024wannier}.
Further differences of the algorithms include the starting point and the partial charge estimate for the \gls{pm} localization functional.
Both works use a similar line search strategy, based on a polynomial approximation approach.~\cite{Abrudan2009}

\part[II]{}

%%%%%%%%%%%%%%%%%%%%%%%%%%%%%%%%%%%%%%%%%%%%%%%%%%%%%%
%%%%%%%%%%%%%%%%%%%%%%%%%%%%%%%%%%%%%%%%%%%%%%%%%%%%%%
%      Properties
%%%%%%%%%%%%%%%%%%%%%%%%%%%%%%%%%%%%%%%%%%%%%%%%%%%%%%
%%%%%%%%%%%%%%%%%%%%%%%%%%%%%%%%%%%%%%%%%%%%%%%%%%%%%%
\section{Properties of Intrinsic Bond Orbitals} \label{sec:prop}

\begin{table*}
    \centering
    \begin{tabular}{cccccccc}
    \hline\hline
 & structure & bond & $a [\AA]$ &  $d_\text{NN} [\AA]$ & $\#\text{NN}$ & $\sigma_\text{HF} [\AA]$ & $\gamma_\text{exp} [eV]$\\
 \hline
C (Diamond) & A4 & covalent & 3.57 &  1.54 & 4 & 0.832 & 5.48 \\
Si & A4 & covalent & 5.431 &  2.35 & 4 & 1.268 & 1.17 \\
Ge & A4 & covalent & 5.652 &  2.45 & 4 & 1.351 & 0.74\\
NaCl & B1 & ionic & 5.569 &  2.78 & 6 & 0.813 & 9.50\\
MgO & B1 & ionic & 4.189 &  2.09 & 6 & 0.817 & 7.22\\
LiF & B1 & ionic & 3.972 &  1.99 & 6 & 0.676 & 14.5\\
SiC & B3 & polar covalent & 4.346 &  1.88 & 4 & 1.009  & 2.42\\
BN & B3 & polar covalent & 3.592 &  1.56 & 4 & 0.806  & 6.22\\
AlP & B3 & polar covalent & 5.451 &  2.36 & 4 & 1.204 & 2.51 \\
GaAs & B3 & polar covalent & 5.64 &  2.44 & 4 & 1.307  & 1.52\\
GaN & B3 & polar covalent & 4.509 &  1.95 & 4 & 0.966 & 3.30 \\
C (Lonsdaleite) & B4 & covalent & 4.347 &  1.54 & 4 & 0.835 & 3.35\\
\hline\hline
    \end{tabular}
    \caption{List of considered materials in Sec. \ref{sec:prop}. The structure column refers to the Strukturbericht designation. The lattice constant $a$, the nearest neighbor distance $d_\text{NN}$, the number of nearest neighbors $\#\text{NN}$, the \gls{ibo} orbital spread $\sigma_\text{HF}$, and the experimental band gap $\gamma_\text{exp}$ are also provided. }
    \label{tab:set}
\end{table*}

\begin{figure*}
    \centering
    \includegraphics[width=1\textwidth]{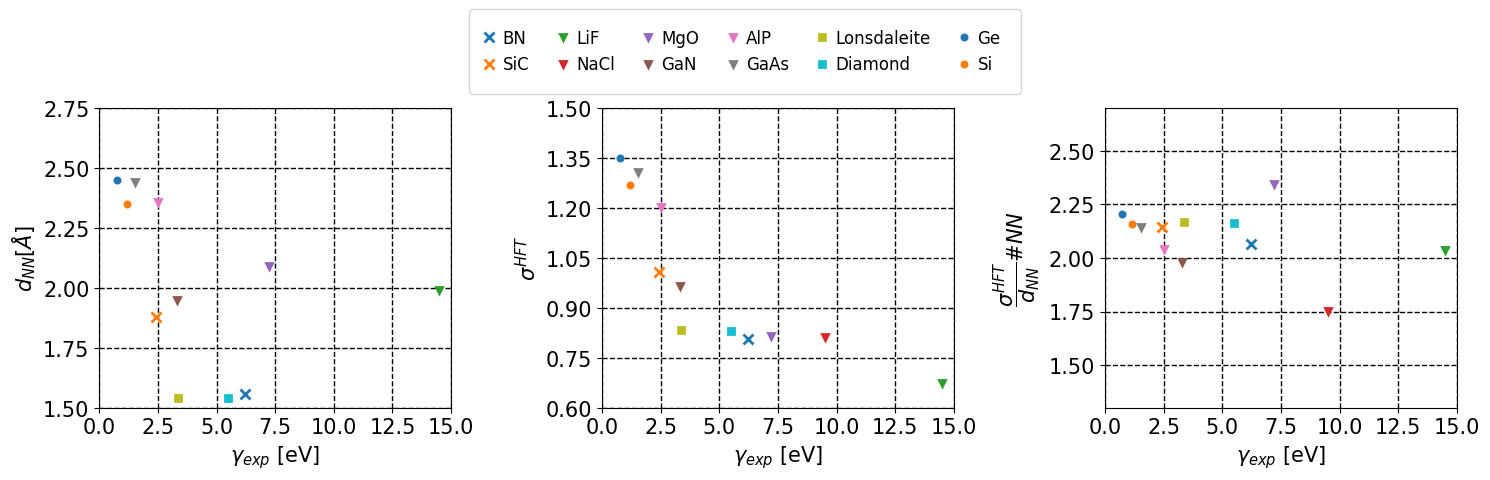}
    \caption{Plot of the nearest neighbor distance $d_\text{NN}$ (left), the \gls{ibo} orbital spread $\sigma_\text{HF}$ (centre), and relation between orbitals spread and $d_\text{NN}/\#\text{NN}$ (right) plotted against the experimental band gap of all considered materials.}
    \label{fig:spreads}
\end{figure*}

In the second main part of the paper, we investigate spatial properties of \glspl{ibo} for a variety of insulating solids, as listed in Tab.~\ref{tab:set}. These \glspl{ibo} were constructed from periodic Hartree-Fock (HF) orbitals, and their spatial character was analyzed in relation to the experimental lattice constant and band gap. A focus was the average orbital spread of the \glspl{ibo} and its relationship with geometric properties, specifically the nearest neighbor distance ($d_\text{NN}$) and the number of nearest neighbors ($\#\text{NN}$). Our findings reveal that the orbital spread of the valence electrons per nearest neighbor distance, multiplied by the number of nearest neighbors, remains relatively stable across the materials considered. This can be captured by the empirical relation $\sigma_{\text{HF}} \approx 2.1 \alpha$, where $\alpha = \frac{d_\text{NN}}{\#\text{NN}}$. This trend was consistent across various crystal structures, independent of the band gap, providing a useful estimate for predicting the spatial extent of localized orbitals (see Fig. \ref{fig:spreads}).

\begin{figure}
\centering
\includegraphics[width=0.4\textwidth]{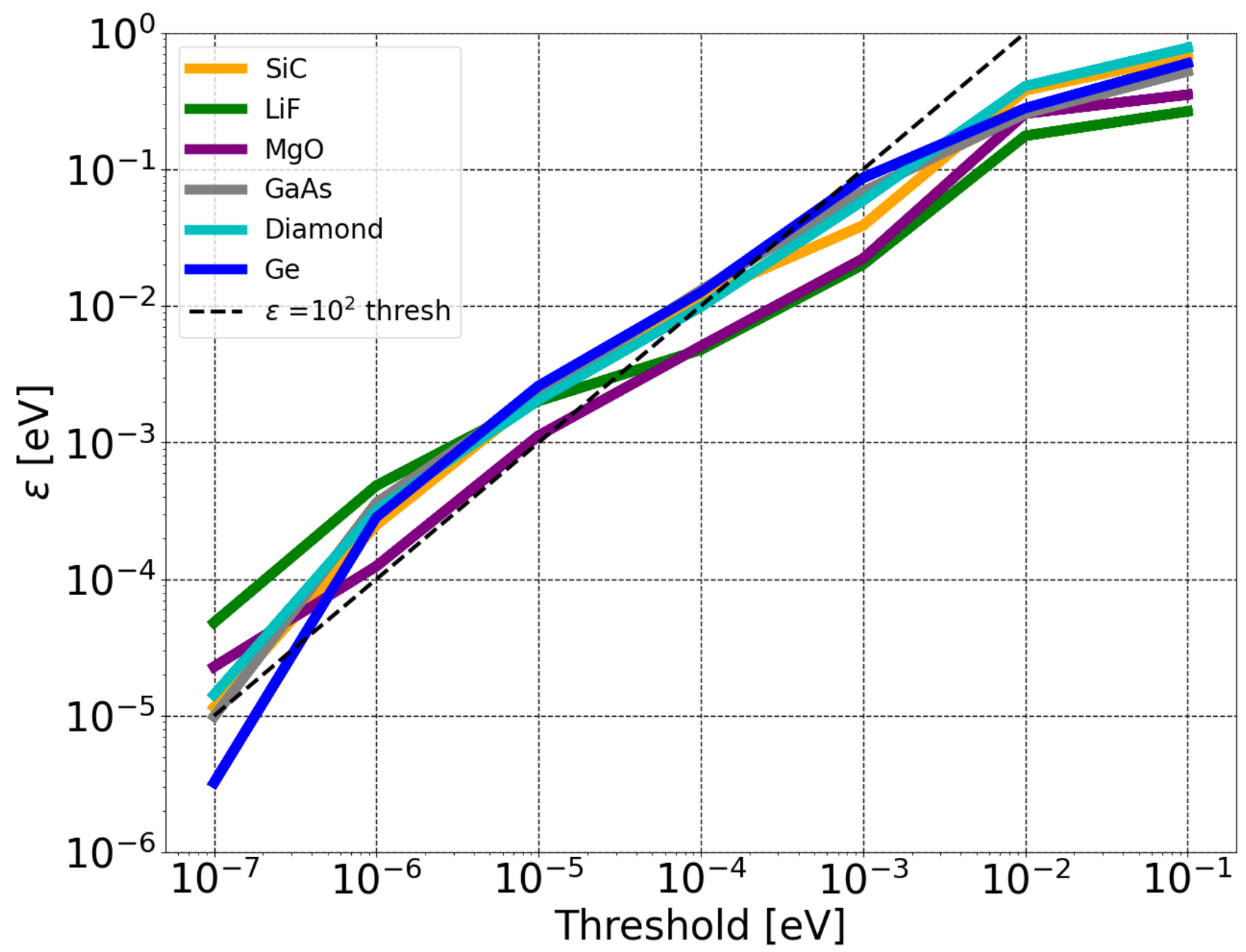}

\caption{Error of the Fock exchange energy per atom $\varepsilon$ for different thresholds using the magnitude cutoff.}
\label{Fig:Error_Methods_1}
\end{figure}

\begin{figure}
\centering
\includegraphics[width=0.41\textwidth]{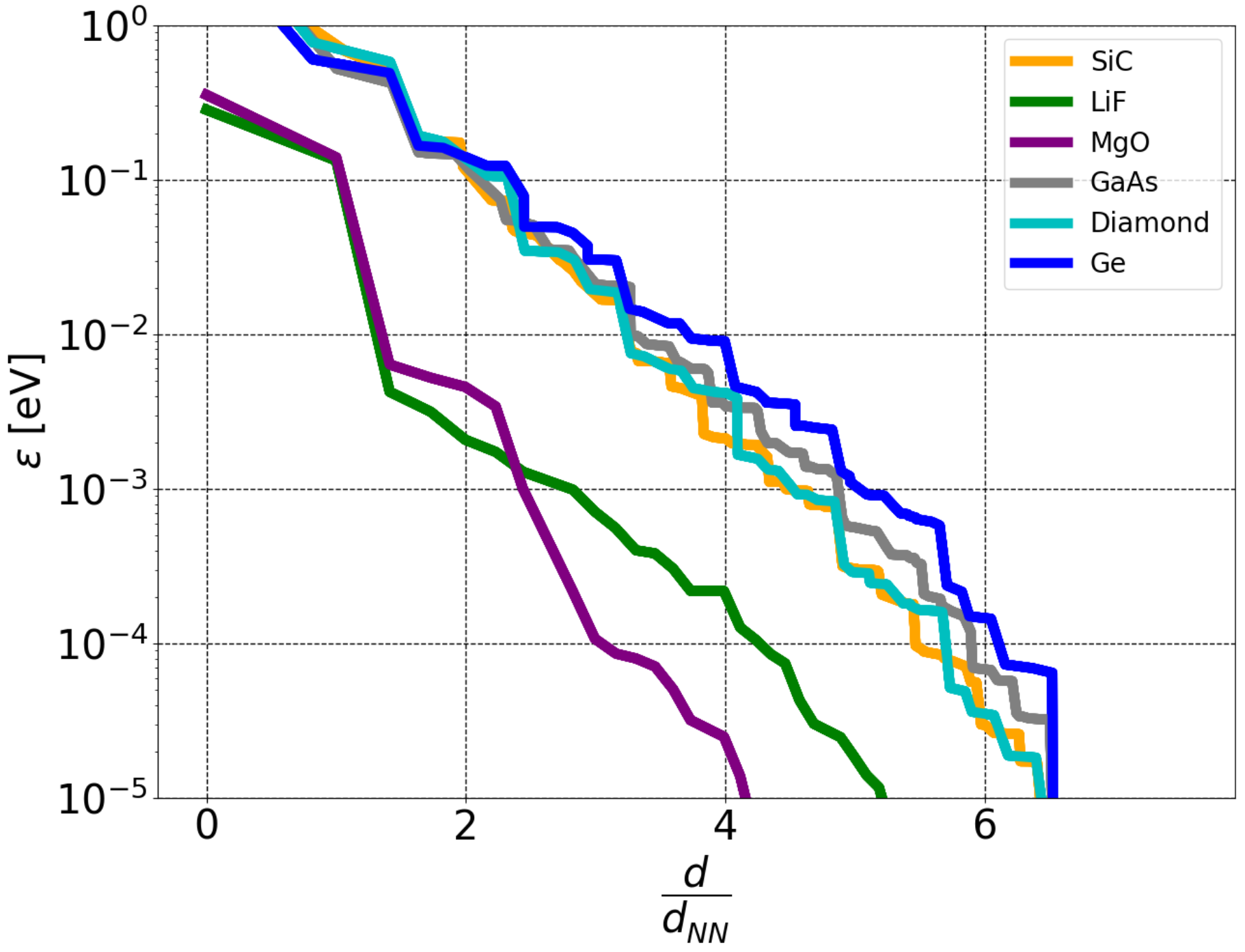}

\caption{Error of the Fock exchange energy per atom with respect to the largest distance between the Wannier orbitals $d_{\text{max}}$ per nearest neighbour distance, using the distance cutoff.}
\label{Fig:Error_Methods_2}
\end{figure}

We also analyzed the decay of the Fock exchange matrix entries,
\begin{equation}
K_{ij} = - \frac{1}{2}  \int \D^3 r_1 \int \D^3 r_2  
\frac{\mathcal W^*_{i}(\bm r_1) \mathcal W_{j}(\bm r_1) \; \mathcal W^*_{j}(\bm r_2)  \mathcal W_{i}(\bm r_2)  }{|\bm r_1 - \bm r_2|} \;,
\end{equation}
in the basis of Wannier orbitals $\mathcal W_{i}(\bm r)$ in the form of \glspl{ibo}. 
To this end, we define the error of the Fock exchange energy per atom as  
\begin{equation}
\varepsilon = \left| \sum_{ij} K_{ij} -  \sum_{ij}^{\text{trunc.}} K_{ij} \right| / N_A\;,
\end{equation}
where $N_A$ is the number of atoms.
Two truncation methods were employed to estimate the error per atom, a magnitude cutoff and a distance cutoff.
The magnitude cutoff eliminates matrix elements below a certain energy threshold (Fig. ~\ref{Fig:Error_Methods_1}), while the distance cutoff uses the inter-orbital distances of the centers of the \glspl{ibo} to determine which elements to retain (Fig.~\ref{Fig:Error_Methods_2}).

It is noteworthy that the magnitude cutoff method yields remarkably consistent error estimates across all considered materials. 
For example, comparing germanium, a narrow-gap semiconductor with an experimental band gap of 
$0.74\,\text{eV}$ and high relative permittivity, to diamond, a wide-gap insulator with a band gap of $5.48\,\text{eV}$ and significantly lower polarizability, reveals no substantial variation in error behavior.
However, the distance cutoff method exhibits a distinct dependence for ionic compounds. 
In these materials, the valence charge is primarily localized on the anions, resulting in a depletion of valence electron density around the cations. 
Consequently, the inter-orbital distance between neighboring bonding orbitals increases to approximately $2d_\text{NN}$.
This increased separation leads to a more rapid decay of the error when employing the distance cutoff for ionic systems.

\begin{figure*}
    \centering
    \includegraphics[width=1\textwidth]{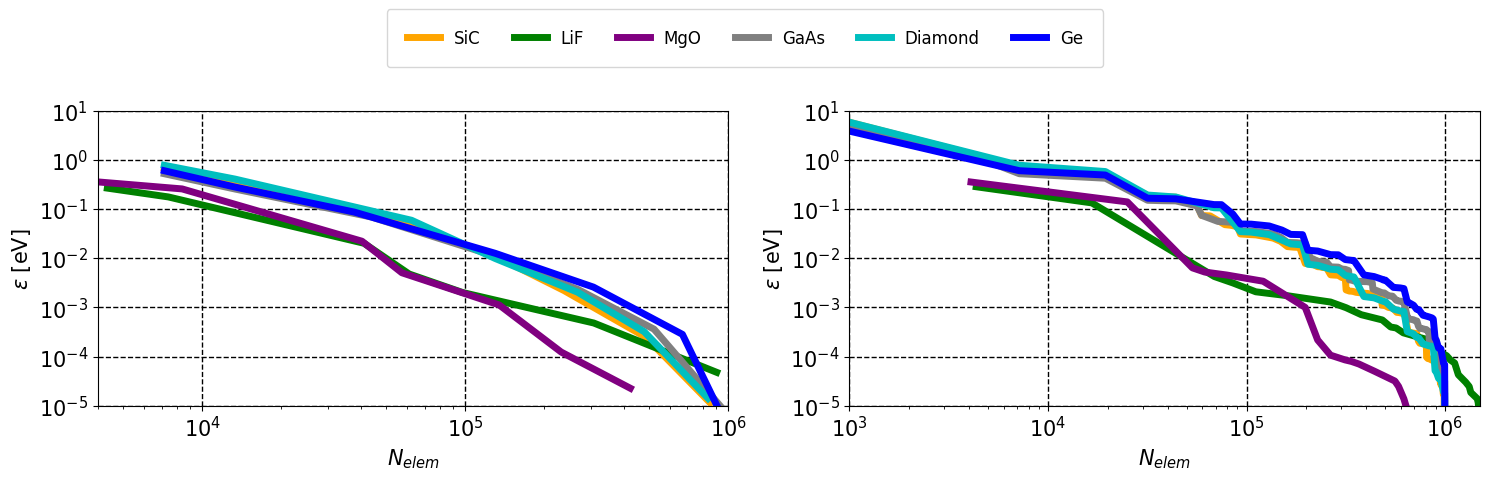}
    \caption{Fock exchange energy error per atom $\varepsilon$ with respect to the number of non-zero elements $N_\text{elem}$ of the Fock exchange matrix using the magnitude cutoff (left) and distance cutoff (right), as explained in the text. For our calculations the full Fock exchange matrix contains $1280^2 \approx 1.6\cdot 10^6$ matrix elements for LiF and $1024^2 \approx 10^6$ for all other materials in the plot. This corresponds to $4\times 4 \times 4$ supercells of the conventional cells.}
    \label{fig:sparsmatr} 
\end{figure*}

Our findings on the truncation of the Fock exchange matrix further support the potential for computational savings due to its sparsity. This holds for both large-gap and small-gap materials, suggesting that the band gap has a relatively minor effect on the decay of matrix elements for practical purposes. 
%However, for applications requiring high accuracy, materials with a small-gap like germanium, show a slightly slower decay of the orbitals and larger errors may accumulate if the truncation is too aggressive.
Figure \ref{fig:sparsmatr} shows the number of remaining non-zero Fock exchange matrix elements in dependence of $\varepsilon$ for both methods.

In summary, the analysis of IBO properties across various bulk 3D materials, exhibiting covalent bonds, polar covalent bonds, and ionic bonds, reveals a correlation between orbital spread and geometric factors. This relationship provides a straightforward way to estimate the spatial extent of localized orbitals, which is crucial for the development of reduced-cost methods. Additionally, the truncation of the Fock exchange matrix demonstrates significant potential for reducing computational costs in large-scale simulations, with manageable errors across a wide range of materials.

%%%%%%%%%%%%%%%%%%%%%%%%%%%%%%%%%%%%%%%%%%%%%%%%%%%%%%
%%%%%%%%%%%%%%%%%%%%%%%%%%%%%%%%%%%%%%%%%%%%%%%%%%%%%%
%      Conclusion
%%%%%%%%%%%%%%%%%%%%%%%%%%%%%%%%%%%%%%%%%%%%%%%%%%%%%%
%%%%%%%%%%%%%%%%%%%%%%%%%%%%%%%%%%%%%%%%%%%%%%%%%%%%%%
\section{Conclusion}

In this work we studied the numerical construction and spatial properties of \glspl{ibo} based on \gls{hf} orbitals for a set of insulating solids.
We reported a relation between the orbital spread measured in units of the nearest neighbor distance and the number of nearest neighbors. 
Independent of the band gap, this relation is relatively stable for all considered 3D semiconductors and insulators.
It suggests that local methods based on the sparsity of Coulomb integrals can also be applied to materials with small band gaps without losing the sparsity.  
We verified this hypothesis for the particular case of the sparsity of the Fock exchange matrix in the basis of \glspl{ibo}.
Whether this can be extended to metallic solids or scenarios involving localized unoccupied orbitals, essential for many-electron correlation methods, remains an open question.
This warrants further investigation in future work.

Additionally, we benchmarked various solvers to optimize the unitary matrix that transforms delocalized Bloch orbitals into localized Wannier orbitals, specifically in the form of \glspl{ibo}. 
The solvers have been implemented within \gls{vasp} and as a standalone open-source software package Lucon.jl~\cite{lucon}.
Our focus was on large simulation cells, which are crucial for realistic models, such as those involving surface phenomena and defects. 
Contrary to a previous study~\cite{Clement2021}, we did not observe a clear performance advantage of the \gls{lbfgs} solver, instead finding that both the \gls{cg} and \gls{lbfgs} solvers exhibited similar performance.
When using stochastic (unbiased) starting points, we found that constructing localized orbitals in supercells of metal oxides pose a significant challenge, requiring an order of magnitude more iteration steps than for the other materials considered.
These findings underscore the importance of optimized initial guesses, the potential of effective preconditioning strategies, and the exploration of non-iterative approaches for efficient Wannier orbital construction in solids.

\section*{Acknowledgements}

T.S. acknowledges support from the Austrian Science Fund
(FWF) [10.55776/ESP335].
The computational results presented have been achieved in part using the Vienna Scientific Cluster (VSC).

\bibliography{refs}

\end{document}

\author{Benjamin W\"ockinger}
\author{Alexander Rumpf}
\author{Tobias Sch\"afer}
\email{tobias.schaefer@tuwien.ac.at}
\affiliation{Institute for Theoretical Physics, TU Wien, Wiedner Hauptstraße 8-10/136, A-1040 Vienna, Austria}

\title{Supplementary information for:\\Exploring the Convergence and Properties of Intrinsic Bond Orbitals in Solids}

\maketitle

\section*{Atomic structures}

Here we provide the considered atomic structures and cells of 
\begin{itemize}
    \item caffeine,
    \item benzene,
    \item coronene,
    \item graphene with flower defect,
    \item silicon with interstitial defect,
\end{itemize}
in the form of \texttt{POSCAR} files for VASP.

\subsection*{Caffeine}
\begin{verbatim}
caffeine molecule
1.0
       19.0000000000   0.0000000000   0.0000000000
        0.0000000000  18.0000000000   0.0000000000
        0.0000000000   0.0000000000  12.0000000000
        O N C H
        2 4 8 10
Cartesian
   0.4700    2.5688    0.0006
  -3.1271   -0.4436   -0.0003
  -0.9686   -1.3125    0.0000
   2.2182    0.1412   -0.0003
  -1.3477    1.0797   -0.0001
   1.4119   -1.9372    0.0002
   0.8579    0.2592   -0.0008
   0.3897   -1.0264   -0.0004
   0.0307    1.4220   -0.0006
  -1.9061   -0.2495   -0.0004
   2.5032   -1.1998    0.0003
  -1.4276   -2.6960    0.0008
   3.1926    1.2061    0.0003
  -2.2969    2.1881    0.0007
   3.5163   -1.5787    0.0008
  -1.0451   -3.1973   -0.8937
  -2.5186   -2.7596    0.0011
  -1.0447   -3.1963    0.8957
   4.1992    0.7801    0.0002
   3.0468    1.8092   -0.8992
   3.0466    1.8083    0.9004
  -1.8087    3.1651   -0.0003
  -2.9322    2.1027    0.8881
  -2.9346    2.1021   -0.8849

\end{verbatim}

\subsection*{Benzene}
\begin{verbatim}
benzene molecule
1.0
  16.2000000000   0.0000000000   0.0000000000
   0.0000000000  16.1000000000   0.0000000000
   0.0000000000   0.0000000000  16.0000000000
    C    H
    6    6
Cart
 8.712645325   9.120995701   8.060540783
 9.258235731   7.840748099   8.124233518
 8.426884963   6.725473338   8.042650428
 7.050422155   6.889925942   7.899686405
 6.504474360   8.171050561   7.838453982
 7.336172399   9.286642887   7.916598567
 9.357841649   9.986399167   8.127737169
10.324954277   7.712900122   8.246743035
 8.850444649   5.731567320   8.094749952
 6.405544304   6.023726297   7.836286520
 5.436217209   8.299221149   7.726296885
 6.913099303  10.281000204   7.867113860
\end{verbatim}

\subsection*{Coronene}
\begin{verbatim}
coronene molecule
1.0
  20.21             0.0000000000   0.0000000000
   0.0000000000  20.21             0.0000000000
   0.0000000000   0.0000000000  14.458
    C    H
   24   12
Cart
 0.06546813853317   -1.23154341226572    1.72900013164781
 1.48753197971382   -1.23154341226572    1.72900013164781
 2.19856390030415    0.00000000000000    1.72900013164781
 1.48753197971382    1.23154341226572    1.72900013164781
 0.06546813853317    1.23154341226572    1.72900013164781
-0.64556378205716    0.00000000000000    1.72900013164781
 2.19529218179055   -2.45742004183783    1.72900013164781
 1.45962227520803   -3.67082275092130    1.72900013164781
 0.09337784303896   -3.67082275092130    1.72900013164781
-0.64229206354356   -2.45742004183783    1.72900013164781
-2.06108418621062    0.00000000000000    1.72900013164781
-2.74408680400649   -1.24381018244192    1.72900013164781
-2.06096458792196   -2.42701256847938    1.72900013164781
-2.74408680400649    1.24381018244192    1.72900013164781
-2.06096458792196    2.42701256847938    1.72900013164781
-0.64229206354356    2.45742004183783    1.72900013164781
 0.09337784303896    3.67082275092130    1.72900013164781
 1.45962227520803    3.67082275092130    1.72900013164781
 2.19529218179055    2.45742004183783    1.72900013164781
 3.61396470616894    2.42701256847938    1.72900013164781
 4.29708692225348    1.24381018244192    1.72900013164781
 3.61408430445760    0.00000000000000    1.72900013164781
 4.29708692225348   -1.24381018244192    1.72900013164781
 3.61396470616894   -2.42701256847938    1.72900013164781
 2.00258058359503   -4.60741684241786    1.72900013164781
-0.44958046534805   -4.60741684241786    1.72900013164781
-2.60059970999888   -3.36552530248663    1.72900013164781
-3.82668023447042   -1.24189153993123    1.72900013164781
-3.82668023447042    1.24189153993123    1.72900013164781
-2.60059970999888    3.36552530248663    1.72900013164781
-0.44958046534805    4.60741684241786    1.72900013164781
 2.00258058359503    4.60741684241786    1.72900013164781
 4.15359982824587    3.36552530248663    1.72900013164781
 5.37968035271741    1.24189153993123    1.72900013164781
 5.37968035271741   -1.24189153993123    1.72900013164781
 4.15359982824587   -3.36552530248663    1.72900013164781
\end{verbatim}

\subsection*{Graphene flower defect}
\begin{verbatim}
graphene flower defect 
  1.0
    22.149000000  0.000000000  0.000000000  
   -11.074500000 19.181596668  0.000000000  
     0.000000000  0.000000000 14.000000000  
  C
  162
  Direct
     0.957441843  0.015639991  0.500000000
     0.021364625  0.079562773  0.500000000
     0.021364625  0.144142492  0.500000000
     0.085114392  0.206615469  0.500000000
     0.943254924  0.199760769  0.500000000
     0.983649019  0.272959127  0.500000000
     0.947107998  0.311418992  0.500000000
     0.984257301  0.385664047  0.500000000
     0.947438566  0.422970210  0.500000000
     0.984436143  0.497175750  0.500000000
     0.947438566  0.534383713  0.500000000
     0.984257301  0.608508611  0.500000000
     0.947107998  0.645604364  0.500000000
     0.983649019  0.720605250  0.500000000
     0.943254924  0.753409513  0.500000000
     0.894138914  0.824664513  0.500000000
     0.892862124  0.887137490  0.500000000
     0.957441843  0.951717209  0.500000000
     0.021364625  0.951717209  0.500000000
     0.085287407  0.015639991  0.500000000
     0.085287407  0.079562773  0.500000000
     0.149867126  0.144142492  0.500000000
     0.148590336  0.206615469  0.500000000
     0.099474326  0.277870470  0.500000000
     0.059080231  0.310674733  0.500000000
     0.095621252  0.385675618  0.500000000
     0.058471949  0.422771371  0.500000000
     0.095290684  0.496896269  0.500000000
     0.058293107  0.534104232  0.500000000
     0.095290684  0.608309772  0.500000000
     0.058471949  0.645615935  0.500000000
     0.095621252  0.719860991  0.500000000
     0.059080231  0.758320855  0.500000000
     0.099474326  0.831519213  0.500000000
     0.957614858  0.824664513  0.500000000
     0.021364625  0.887137490  0.500000000
     0.085114392  0.888414280  0.500000000
     0.148590336  0.951890224  0.500000000
     0.149867126  0.015639991  0.500000000
     0.212340103  0.079389758  0.500000000
     0.212340103  0.142865702  0.500000000
     0.205485403  0.277870470  0.500000000
     0.172811607  0.318533955  0.500000000
     0.208734293  0.390379327  0.500000000
     0.169812852  0.424797347  0.500000000
     0.206784260  0.498318887  0.500000000
     0.169509588  0.534538478  0.500000000
     0.206634669  0.608275013  0.500000000
     0.169509588  0.644886466  0.500000000
     0.206784260  0.718380731  0.500000000
     0.169812852  0.754930862  0.500000000
     0.208734293  0.828270323  0.500000000
     0.172811607  0.864193009  0.500000000
     0.205485403  0.937530290  0.500000000
     0.278683761  0.977924385  0.500000000
     0.316399367  0.053355597  0.500000000
     0.283595104  0.093749692  0.500000000
     0.324258589  0.167086973  0.500000000
     0.283595104  0.199760769  0.500000000
     0.316399367  0.272959127  0.500000000
     0.278683761  0.310674733  0.500000000
     0.317143626  0.385675618  0.500000000
     0.282073754  0.424797347  0.500000000
     0.318623886  0.498318887  0.500000000
     0.280958778  0.534828298  0.500000000
     0.317782651  0.608476043  0.500000000
     0.280638437  0.645276897  0.500000000
     0.317782651  0.719221965  0.500000000
     0.280958778  0.756045838  0.500000000
     0.318623886  0.830220356  0.500000000
     0.282073754  0.867191765  0.500000000
     0.317143626  0.941383364  0.500000000
     0.391388681  0.978532667  0.500000000
     0.428496005  0.052747315  0.500000000
     0.391400252  0.089896618  0.500000000
     0.430521981  0.164088218  0.500000000
     0.396103961  0.203009659  0.500000000
     0.430521981  0.276349120  0.500000000
     0.391400252  0.311418992  0.500000000
     0.428496005  0.385664047  0.500000000
     0.391388681  0.422771371  0.500000000
     0.428694844  0.496896269  0.500000000
     0.392118150  0.534538478  0.500000000
     0.428729603  0.608275013  0.500000000
     0.391727719  0.645276897  0.500000000
     0.428528573  0.719221965  0.500000000
     0.391727719  0.756366179  0.500000000
     0.428729603  0.830369947  0.500000000
     0.392118150  0.867495028  0.500000000
     0.428694844  0.941713932  0.500000000
     0.502900384  0.978711509  0.500000000
     0.539828866  0.052568473  0.500000000
     0.502620903  0.089566050  0.500000000
     0.540263112  0.163784954  0.500000000
     0.504043521  0.201059626  0.500000000
     0.540552932  0.275234144  0.500000000
     0.504043521  0.312899252  0.500000000
     0.540263112  0.386393516  0.500000000
     0.502620903  0.422970210  0.500000000
     0.539828866  0.497175750  0.500000000
     0.502900384  0.534104232  0.500000000
     0.540108347  0.608309772  0.500000000
     0.502466138  0.644886466  0.500000000
     0.538685729  0.718380731  0.500000000
     0.502176318  0.756045838  0.500000000
     0.538685729  0.830220356  0.500000000
     0.502466138  0.867495028  0.500000000
     0.540108347  0.941713932  0.500000000
     0.614233245  0.978532667  0.500000000
     0.651340569  0.052747315  0.500000000
     0.614034406  0.089566050  0.500000000
     0.650611100  0.163784954  0.500000000
     0.613999647  0.200910035  0.500000000
     0.651001531  0.274913803  0.500000000
     0.614200677  0.312058017  0.500000000
     0.651001531  0.386003085  0.500000000
     0.613999647  0.423004969  0.500000000
     0.650611100  0.496741504  0.500000000
     0.614034406  0.534383713  0.500000000
     0.651340569  0.608508611  0.500000000
     0.614233245  0.645615935  0.500000000
     0.651328998  0.719860991  0.500000000
     0.612207270  0.754930862  0.500000000
     0.646625289  0.828270323  0.500000000
     0.612207270  0.867191765  0.500000000
     0.651328998  0.941383364  0.500000000
     0.726329884  0.977924385  0.500000000
     0.764045489  0.053355597  0.500000000
     0.725585625  0.089896618  0.500000000
     0.760655496  0.164088218  0.500000000
     0.724105365  0.201059626  0.500000000
     0.761770472  0.275234144  0.500000000
     0.724946599  0.312058017  0.500000000
     0.762090813  0.386003085  0.500000000
     0.724946599  0.422803939  0.500000000
     0.761770472  0.496451684  0.500000000
     0.724105365  0.532961095  0.500000000
     0.760655496  0.606482635  0.500000000
     0.725585625  0.645604364  0.500000000
     0.764045489  0.720605250  0.500000000
     0.726329884  0.758320855  0.500000000
     0.759134147  0.831519213  0.500000000
     0.718470661  0.864193009  0.500000000
     0.759134147  0.937530290  0.500000000
     0.894138914  0.079389758  0.500000000
     0.957614858  0.142865702  0.500000000
     0.837243847  0.093749692  0.500000000
     0.869917643  0.167086973  0.500000000
     0.833994957  0.203009659  0.500000000
     0.872916399  0.276349120  0.500000000
     0.835944990  0.312899252  0.500000000
     0.873219663  0.386393516  0.500000000
     0.836094581  0.423004969  0.500000000
     0.873219663  0.496741504  0.500000000
     0.835944990  0.532961095  0.500000000
     0.872916399  0.606482635  0.500000000
     0.833994957  0.640900655  0.500000000
     0.869917643  0.712746027  0.500000000
     0.837243847  0.753409513  0.500000000
     0.830389147  0.888414280  0.500000000
     0.830389147  0.951890224  0.500000000
     0.892862124  0.015639991  0.500000000
\end{verbatim}

\subsection*{Silicon X interstitial}
\begin{verbatim}
Silicon X interstitial (520 atoms)
1.0
       21.8788528442         0.0000000000         0.0000000000
        0.0000000000        21.8788528442         0.0000000000
        0.0000000000         0.0000000000        21.8788528442
   Si
  520
Direct
     0.101965073         0.022856597         0.027526432
     0.101965073         0.022856597         0.527526454
     0.101965073         0.522856605         0.027526432
     0.101965073         0.522856605         0.527526454
     0.601965073         0.022856597         0.027526432
     0.601965073         0.022856597         0.527526454
     0.601965073         0.522856605         0.027526432
     0.601965073         0.522856605         0.527526454
     0.022856597         0.101965073         0.027526432
     0.022856597         0.101965073         0.527526454
     0.022856597         0.601965073         0.027526432
     0.022856597         0.601965073         0.527526454
     0.522856605         0.101965073         0.027526432
     0.522856605         0.101965073         0.527526454
     0.522856605         0.601965073         0.027526432
     0.522856605         0.601965073         0.527526454
     0.995866032         0.995866032         0.996017547
     0.995866032         0.995866032         0.496017460
     0.995866032         0.495866076         0.996017547
     0.995866032         0.495866076         0.496017460
     0.495866076         0.995866032         0.996017547
     0.495866076         0.995866032         0.496017460
     0.495866076         0.495866076         0.996017547
     0.495866076         0.495866076         0.496017460
     0.250261664         0.998747605         0.000861127
     0.250261664         0.998747605         0.500861098
     0.250261664         0.498747605         0.000861127
     0.250261664         0.498747605         0.500861098
     0.750261708         0.998747605         0.000861127
     0.750261708         0.998747605         0.500861098
     0.750261708         0.498747605         0.000861127
     0.750261708         0.498747605         0.500861098
     0.998747605         0.250261664         0.000861127
     0.998747605         0.250261664         0.500861098
     0.998747605         0.750261708         0.000861127
     0.998747605         0.750261708         0.500861098
     0.498747605         0.250261664         0.000861127
     0.498747605         0.250261664         0.500861098
     0.498747605         0.750261708         0.000861127
     0.498747605         0.750261708         0.500861098
     0.249773512         0.249773512         0.999446073
     0.249773512         0.249773512         0.499446160
     0.249773512         0.749773512         0.999446073
     0.249773512         0.749773512         0.499446160
     0.749773512         0.249773512         0.999446073
     0.749773512         0.249773512         0.499446160
     0.749773512         0.749773512         0.999446073
     0.749773512         0.749773512         0.499446160
     0.999731841         0.999731841         0.250053876
     0.999731841         0.999731841         0.750053876
     0.999731841         0.499731798         0.250053876
     0.999731841         0.499731798         0.750053876
     0.499731798         0.999731841         0.250053876
     0.499731798         0.999731841         0.750053876
     0.499731798         0.499731798         0.250053876
     0.499731798         0.499731798         0.750053876
     0.249916397         0.999819716         0.249595888
     0.249916397         0.999819716         0.749595844
     0.249916397         0.499819760         0.249595888
     0.249916397         0.499819760         0.749595844
     0.749916397         0.999819716         0.249595888
     0.749916397         0.999819716         0.749595844
     0.749916397         0.499819760         0.249595888
     0.749916397         0.499819760         0.749595844
     0.999819716         0.249916397         0.249595888
     0.999819716         0.249916397         0.749595844
     0.999819716         0.749916397         0.249595888
     0.999819716         0.749916397         0.749595844
     0.499819760         0.249916397         0.249595888
     0.499819760         0.249916397         0.749595844
     0.499819760         0.749916397         0.249595888
     0.499819760         0.749916397         0.749595844
     0.249601772         0.249601772         0.250873303
     0.249601772         0.249601772         0.750873346
     0.249601772         0.749601772         0.250873303
     0.249601772         0.749601772         0.750873346
     0.749601772         0.249601772         0.250873303
     0.749601772         0.249601772         0.750873346
     0.749601772         0.749601772         0.250873303
     0.749601772         0.749601772         0.750873346
     0.128970455         0.128970455         0.996066454
     0.128970455         0.128970455         0.496066454
     0.128970455         0.628970466         0.996066454
     0.128970455         0.628970466         0.496066454
     0.628970466         0.128970455         0.996066454
     0.628970466         0.128970455         0.496066454
     0.628970466         0.628970466         0.996066454
     0.628970466         0.628970466         0.496066454
     0.374691565         0.126273710         0.000893315
     0.374691565         0.126273710         0.500893310
     0.374691565         0.626273710         0.000893315
     0.374691565         0.626273710         0.500893310
     0.874691522         0.126273710         0.000893315
     0.874691522         0.126273710         0.500893310
     0.874691522         0.626273710         0.000893315
     0.874691522         0.626273710         0.500893310
     0.126273710         0.374691565         0.000893315
     0.126273710         0.374691565         0.500893310
     0.126273710         0.874691522         0.000893315
     0.126273710         0.874691522         0.500893310
     0.626273710         0.374691565         0.000893315
     0.626273710         0.374691565         0.500893310
     0.626273710         0.874691522         0.000893315
     0.626273710         0.874691522         0.500893310
     0.375103262         0.375103262         0.999455139
     0.375103262         0.375103262         0.499455226
     0.375103262         0.875103262         0.999455139
     0.375103262         0.875103262         0.499455226
     0.875103262         0.375103262         0.999455139
     0.875103262         0.375103262         0.499455226
     0.875103262         0.875103262         0.999455139
     0.875103262         0.875103262         0.499455226
     0.125214533         0.125214533         0.250042739
     0.125214533         0.125214533         0.750042717
     0.125214533         0.625214501         0.250042739
     0.125214533         0.625214501         0.750042717
     0.625214501         0.125214533         0.250042739
     0.625214501         0.125214533         0.750042717
     0.625214501         0.625214501         0.250042739
     0.625214501         0.625214501         0.750042717
     0.374968878         0.125094337         0.249609880
     0.374968878         0.125094337         0.749609880
     0.374968878         0.625094370         0.249609880
     0.374968878         0.625094370         0.749609880
     0.874968921         0.125094337         0.249609880
     0.874968921         0.125094337         0.749609880
     0.874968921         0.625094370         0.249609880
     0.874968921         0.625094370         0.749609880
     0.125094337         0.374968878         0.249609880
     0.125094337         0.374968878         0.749609880
     0.125094337         0.874968921         0.249609880
     0.125094337         0.874968921         0.749609880
     0.625094370         0.374968878         0.249609880
     0.625094370         0.374968878         0.749609880
     0.625094370         0.874968921         0.249609880
     0.625094370         0.874968921         0.749609880
     0.375383408         0.375383408         0.250858744
     0.375383408         0.375383408         0.750858788
     0.375383408         0.875383364         0.250858744
     0.375383408         0.875383364         0.750858788
     0.875383364         0.375383408         0.250858744
     0.875383364         0.375383408         0.750858788
     0.875383364         0.875383364         0.250858744
     0.875383364         0.875383364         0.750858788
     0.996615499         0.128242085         0.127911409
     0.996615499         0.128242085         0.627911387
     0.996615499         0.628242096         0.127911409
     0.996615499         0.628242096         0.627911387
     0.496615543         0.128242085         0.127911409
     0.496615543         0.128242085         0.627911387
     0.496615543         0.628242096         0.127911409
     0.496615543         0.628242096         0.627911387
     0.250542115         0.124734053         0.125272289
     0.250542115         0.124734053         0.625272256
     0.250542115         0.624734064         0.125272289
     0.250542115         0.624734064         0.625272256
     0.750542071         0.124734053         0.125272289
     0.750542071         0.124734053         0.625272256
     0.750542071         0.624734064         0.125272289
     0.750542071         0.624734064         0.625272256
     0.000194865         0.374353621         0.125257240
     0.000194865         0.374353621         0.625257261
     0.000194865         0.874353621         0.125257240
     0.000194865         0.874353621         0.625257261
     0.500194842         0.374353621         0.125257240
     0.500194842         0.374353621         0.625257261
     0.500194842         0.874353621         0.125257240
     0.500194842         0.874353621         0.625257261
     0.252227936         0.372697941         0.125980597
     0.252227936         0.372697941         0.625980619
     0.252227936         0.872697941         0.125980597
     0.252227936         0.872697941         0.625980619
     0.752227914         0.372697941         0.125980597
     0.752227914         0.372697941         0.625980619
     0.752227914         0.872697941         0.125980597
     0.752227914         0.872697941         0.625980619
     0.000023466         0.124920287         0.374031673
     0.000023466         0.124920287         0.874031673
     0.000023466         0.624920320         0.374031673
     0.000023466         0.624920320         0.874031673
     0.500023494         0.124920287         0.374031673
     0.500023494         0.124920287         0.874031673
     0.500023494         0.624920320         0.374031673
     0.500023494         0.624920320         0.874031673
     0.250766227         0.124541423         0.374286930
     0.250766227         0.124541423         0.874286930
     0.250766227         0.624541402         0.374286930
     0.250766227         0.624541402         0.874286930
     0.750766205         0.124541423         0.374286930
     0.750766205         0.124541423         0.874286930
     0.750766205         0.624541402         0.374286930
     0.750766205         0.624541402         0.874286930
     0.000415239         0.374151281         0.374307242
     0.000415239         0.374151281         0.874307242
     0.000415239         0.874151281         0.374307242
     0.000415239         0.874151281         0.874307242
     0.500415271         0.374151281         0.374307242
     0.500415271         0.374151281         0.874307242
     0.500415271         0.874151281         0.374307242
     0.500415271         0.874151281         0.874307242
     0.249964889         0.374980472         0.375186517
     0.249964889         0.374980472         0.875186604
     0.249964889         0.874980429         0.375186517
     0.249964889         0.874980429         0.875186604
     0.749964867         0.374980472         0.375186517
     0.749964867         0.374980472         0.875186604
     0.749964867         0.874980429         0.375186517
     0.749964867         0.874980429         0.875186604
     0.128242085         0.996615499         0.127911409
     0.128242085         0.996615499         0.627911387
     0.128242085         0.496615543         0.127911409
     0.128242085         0.496615543         0.627911387
     0.628242096         0.996615499         0.127911409
     0.628242096         0.996615499         0.627911387
     0.628242096         0.496615543         0.127911409
     0.628242096         0.496615543         0.627911387
     0.374353621         0.000194865         0.125257240
     0.374353621         0.000194865         0.625257261
     0.374353621         0.500194842         0.125257240
     0.374353621         0.500194842         0.625257261
     0.874353621         0.000194865         0.125257240
     0.874353621         0.000194865         0.625257261
     0.874353621         0.500194842         0.125257240
     0.874353621         0.500194842         0.625257261
     0.124734053         0.250542115         0.125272289
     0.124734053         0.250542115         0.625272256
     0.124734053         0.750542071         0.125272289
     0.124734053         0.750542071         0.625272256
     0.624734064         0.250542115         0.125272289
     0.624734064         0.250542115         0.625272256
     0.624734064         0.750542071         0.125272289
     0.624734064         0.750542071         0.625272256
     0.372697941         0.252227936         0.125980597
     0.372697941         0.252227936         0.625980619
     0.372697941         0.752227914         0.125980597
     0.372697941         0.752227914         0.625980619
     0.872697941         0.252227936         0.125980597
     0.872697941         0.252227936         0.625980619
     0.872697941         0.752227914         0.125980597
     0.872697941         0.752227914         0.625980619
     0.124920287         0.000023466         0.374031673
     0.124920287         0.000023466         0.874031673
     0.124920287         0.500023494         0.374031673
     0.124920287         0.500023494         0.874031673
     0.624920320         0.000023466         0.374031673
     0.624920320         0.000023466         0.874031673
     0.624920320         0.500023494         0.374031673
     0.624920320         0.500023494         0.874031673
     0.374151281         0.000415239         0.374307242
     0.374151281         0.000415239         0.874307242
     0.374151281         0.500415271         0.374307242
     0.374151281         0.500415271         0.874307242
     0.874151281         0.000415239         0.374307242
     0.874151281         0.000415239         0.874307242
     0.874151281         0.500415271         0.374307242
     0.874151281         0.500415271         0.874307242
     0.124541423         0.250766227         0.374286930
     0.124541423         0.250766227         0.874286930
     0.124541423         0.750766205         0.374286930
     0.124541423         0.750766205         0.874286930
     0.624541402         0.250766227         0.374286930
     0.624541402         0.250766227         0.874286930
     0.624541402         0.750766205         0.374286930
     0.624541402         0.750766205         0.874286930
     0.374980472         0.249964889         0.375186517
     0.374980472         0.249964889         0.875186604
     0.374980472         0.749964867         0.375186517
     0.374980472         0.749964867         0.875186604
     0.874980429         0.249964889         0.375186517
     0.874980429         0.249964889         0.875186604
     0.874980429         0.749964867         0.375186517
     0.874980429         0.749964867         0.875186604
     0.312471602         0.062497821         0.062036699
     0.312471602         0.062497821         0.562036672
     0.312471602         0.562497799         0.062036699
     0.312471602         0.562497799         0.562036672
     0.812471645         0.062497821         0.062036699
     0.812471645         0.062497821         0.562036672
     0.812471645         0.562497799         0.062036699
     0.812471645         0.562497799         0.562036672
     0.062497821         0.312471602         0.062036699
     0.062497821         0.312471602         0.562036672
     0.062497821         0.812471645         0.062036699
     0.062497821         0.812471645         0.562036672
     0.562497799         0.312471602         0.062036699
     0.562497799         0.312471602         0.562036672
     0.562497799         0.812471645         0.062036699
     0.562497799         0.812471645         0.562036672
     0.312434486         0.312434486         0.061682208
     0.312434486         0.312434486         0.561682208
     0.312434486         0.812434508         0.061682208
     0.312434486         0.812434508         0.561682208
     0.812434508         0.312434486         0.061682208
     0.812434508         0.312434486         0.561682208
     0.812434508         0.812434508         0.061682208
     0.812434508         0.812434508         0.561682208
     0.062472528         0.062472528         0.311603464
     0.062472528         0.062472528         0.811603442
     0.062472528         0.562472517         0.311603464
     0.062472528         0.562472517         0.811603442
     0.562472517         0.062472528         0.311603464
     0.562472517         0.062472528         0.811603442
     0.562472517         0.562472517         0.311603464
     0.562472517         0.562472517         0.811603442
     0.312437733         0.062452461         0.311262403
     0.312437733         0.062452461         0.811262403
     0.312437733         0.562452423         0.311262403
     0.312437733         0.562452423         0.811262403
     0.812437733         0.062452461         0.311262403
     0.812437733         0.062452461         0.811262403
     0.812437733         0.562452423         0.311262403
     0.812437733         0.562452423         0.811262403
     0.062452461         0.312437733         0.311262403
     0.062452461         0.312437733         0.811262403
     0.062452461         0.812437733         0.311262403
     0.062452461         0.812437733         0.811262403
     0.562452423         0.312437733         0.311262403
     0.562452423         0.312437733         0.811262403
     0.562452423         0.812437733         0.311262403
     0.562452423         0.812437733         0.811262403
     0.312497886         0.312497886         0.312714021
     0.312497886         0.312497886         0.812714000
     0.312497886         0.812497886         0.312714021
     0.312497886         0.812497886         0.812714000
     0.812497886         0.312497886         0.312714021
     0.812497886         0.312497886         0.812714000
     0.812497886         0.812497886         0.312714021
     0.812497886         0.812497886         0.812714000
     0.188178308         0.188178308         0.064115376
     0.188178308         0.188178308         0.564115425
     0.188178308         0.688178308         0.064115376
     0.188178308         0.688178308         0.564115425
     0.688178308         0.188178308         0.064115376
     0.688178308         0.188178308         0.564115425
     0.688178308         0.688178308         0.064115376
     0.688178308         0.688178308         0.564115425
     0.434546005         0.190447871         0.064981503
     0.434546005         0.190447871         0.564981536
     0.434546005         0.690447849         0.064981503
     0.434546005         0.690447849         0.564981536
     0.934546005         0.190447871         0.064981503
     0.934546005         0.190447871         0.564981536
     0.934546005         0.690447849         0.064981503
     0.934546005         0.690447849         0.564981536
     0.190447871         0.434546005         0.064981503
     0.190447871         0.434546005         0.564981536
     0.190447871         0.934546005         0.064981503
     0.190447871         0.934546005         0.564981536
     0.690447849         0.434546005         0.064981503
     0.690447849         0.434546005         0.564981536
     0.690447849         0.934546005         0.064981503
     0.690447849         0.934546005         0.564981536
     0.436692233         0.436692233         0.064126284
     0.436692233         0.436692233         0.564126235
     0.436692233         0.936692233         0.064126284
     0.436692233         0.936692233         0.564126235
     0.936692233         0.436692233         0.064126284
     0.936692233         0.436692233         0.564126235
     0.936692233         0.936692233         0.064126284
     0.936692233         0.936692233         0.564126235
     0.187339331         0.187339331         0.313190666
     0.187339331         0.187339331         0.813190687
     0.187339331         0.687339310         0.313190666
     0.187339331         0.687339310         0.813190687
     0.687339310         0.187339331         0.313190666
     0.687339310         0.187339331         0.813190687
     0.687339310         0.687339310         0.313190666
     0.687339310         0.687339310         0.813190687
     0.437344889         0.187558954         0.312108855
     0.437344889         0.187558954         0.812108899
     0.437344889         0.687558954         0.312108855
     0.437344889         0.687558954         0.812108899
     0.937344933         0.187558954         0.312108855
     0.937344933         0.187558954         0.812108899
     0.937344933         0.687558954         0.312108855
     0.937344933         0.687558954         0.812108899
     0.187558954         0.437344889         0.312108855
     0.187558954         0.437344889         0.812108899
     0.187558954         0.937344933         0.312108855
     0.187558954         0.937344933         0.812108899
     0.687558954         0.437344889         0.312108855
     0.687558954         0.437344889         0.812108899
     0.687558954         0.937344933         0.312108855
     0.687558954         0.937344933         0.812108899
     0.437611391         0.437611391         0.313231683
     0.437611391         0.437611391         0.813231748
     0.437611391         0.937611435         0.313231683
     0.437611391         0.937611435         0.813231748
     0.937611435         0.437611391         0.313231683
     0.937611435         0.437611391         0.813231748
     0.937611435         0.937611435         0.313231683
     0.937611435         0.937611435         0.813231748
     0.063016087         0.187850498         0.187822536
     0.063016087         0.187850498         0.687822536
     0.063016087         0.687850476         0.187822536
     0.063016087         0.687850476         0.687822536
     0.563016070         0.187850498         0.187822536
     0.563016070         0.187850498         0.687822536
     0.563016070         0.687850476         0.187822536
     0.563016070         0.687850476         0.687822536
     0.312011064         0.187853593         0.187964309
     0.312011064         0.187853593         0.687964330
     0.312011064         0.687853571         0.187964309
     0.312011064         0.687853571         0.687964330
     0.812011086         0.187853593         0.187964309
     0.812011086         0.187853593         0.687964330
     0.812011086         0.687853571         0.187964309
     0.812011086         0.687853571         0.687964330
     0.061884247         0.437020065         0.187859041
     0.061884247         0.437020065         0.687859063
     0.061884247         0.937020021         0.187859041
     0.061884247         0.937020021         0.687859063
     0.561884242         0.437020065         0.187859041
     0.561884242         0.437020065         0.687859063
     0.561884242         0.937020021         0.187859041
     0.561884242         0.937020021         0.687859063
     0.312906466         0.437078169         0.187950338
     0.312906466         0.437078169         0.687950295
     0.312906466         0.937078082         0.187950338
     0.312906466         0.937078082         0.687950295
     0.812906488         0.437078169         0.187950338
     0.812906488         0.437078169         0.687950295
     0.812906488         0.937078082         0.187950338
     0.812906488         0.937078082         0.687950295
     0.060695154         0.188620517         0.436959607
     0.060695154         0.188620517         0.936959607
     0.060695154         0.688620517         0.436959607
     0.060695154         0.688620517         0.936959607
     0.560695181         0.188620517         0.436959607
     0.560695181         0.188620517         0.936959607
     0.560695181         0.688620517         0.436959607
     0.560695181         0.688620517         0.936959607
     0.312259673         0.187267235         0.437182346
     0.312259673         0.187267235         0.937182346
     0.312259673         0.687267257         0.437182346
     0.312259673         0.687267257         0.937182346
     0.812259716         0.187267235         0.437182346
     0.812259716         0.187267235         0.937182346
     0.812259716         0.687267257         0.437182346
     0.812259716         0.687267257         0.937182346
     0.064288969         0.436303028         0.436972030
     0.064288969         0.436303028         0.936971986
     0.064288969         0.936302984         0.436972030
     0.064288969         0.936302984         0.936971986
     0.564288952         0.436303028         0.436972030
     0.564288952         0.436303028         0.936971986
     0.564288952         0.936302984         0.436972030
     0.564288952         0.936302984         0.936971986
     0.312683204         0.437666967         0.437191064
     0.312683204         0.437666967         0.937191064
     0.312683204         0.937667054         0.437191064
     0.312683204         0.937667054         0.937191064
     0.812683226         0.437666967         0.437191064
     0.812683226         0.437666967         0.937191064
     0.812683226         0.937667054         0.437191064
     0.812683226         0.937667054         0.937191064
     0.187850498         0.063016087         0.187822536
     0.187850498         0.063016087         0.687822536
     0.187850498         0.563016070         0.187822536
     0.187850498         0.563016070         0.687822536
     0.687850476         0.063016087         0.187822536
     0.687850476         0.063016087         0.687822536
     0.687850476         0.563016070         0.187822536
     0.687850476         0.563016070         0.687822536
     0.437020065         0.061884247         0.187859041
     0.437020065         0.061884247         0.687859063
     0.437020065         0.561884242         0.187859041
     0.437020065         0.561884242         0.687859063
     0.937020021         0.061884247         0.187859041
     0.937020021         0.061884247         0.687859063
     0.937020021         0.561884242         0.187859041
     0.937020021         0.561884242         0.687859063
     0.187853593         0.312011064         0.187964309
     0.187853593         0.312011064         0.687964330
     0.187853593         0.812011086         0.187964309
     0.187853593         0.812011086         0.687964330
     0.687853571         0.312011064         0.187964309
     0.687853571         0.312011064         0.687964330
     0.687853571         0.812011086         0.187964309
     0.687853571         0.812011086         0.687964330
     0.437078169         0.312906466         0.187950338
     0.437078169         0.312906466         0.687950295
     0.437078169         0.812906488         0.187950338
     0.437078169         0.812906488         0.687950295
     0.937078082         0.312906466         0.187950338
     0.937078082         0.312906466         0.687950295
     0.937078082         0.812906488         0.187950338
     0.937078082         0.812906488         0.687950295
     0.188620517         0.060695154         0.436959607
     0.188620517         0.060695154         0.936959607
     0.188620517         0.560695181         0.436959607
     0.188620517         0.560695181         0.936959607
     0.688620517         0.060695154         0.436959607
     0.688620517         0.060695154         0.936959607
     0.688620517         0.560695181         0.436959607
     0.688620517         0.560695181         0.936959607
     0.436303028         0.064288969         0.436972030
     0.436303028         0.064288969         0.936971986
     0.436303028         0.564288952         0.436972030
     0.436303028         0.564288952         0.936971986
     0.936302984         0.064288969         0.436972030
     0.936302984         0.064288969         0.936971986
     0.936302984         0.564288952         0.436972030
     0.936302984         0.564288952         0.936971986
     0.187267235         0.312259673         0.437182346
     0.187267235         0.312259673         0.937182346
     0.187267235         0.812259716         0.437182346
     0.187267235         0.812259716         0.937182346
     0.687267257         0.312259673         0.437182346
     0.687267257         0.312259673         0.937182346
     0.687267257         0.812259716         0.437182346
     0.687267257         0.812259716         0.937182346
     0.437666967         0.312683204         0.437191064
     0.437666967         0.312683204         0.937191064
     0.437666967         0.812683226         0.437191064
     0.437666967         0.812683226         0.937191064
     0.937667054         0.312683204         0.437191064
     0.937667054         0.312683204         0.937191064
     0.937667054         0.812683226         0.437191064
     0.937667054         0.812683226         0.937191064
\end{verbatim}

%\bibliography{refs}% Produces the bibliography via BibTeX.